\documentclass[11pt,a4paper]{article}
\usepackage{jheppub_kim}
\usepackage{rotating}
\usepackage{graphicx,epsfig}
\usepackage{amsmath}
\usepackage {amssymb}
\usepackage{subfigure}

\usepackage{amsfonts}
\usepackage{bm}

\usepackage{relsize}

\begin{document}

\title{Logarithmic-corrected $R^2$ Gravity Inflation in the Presence of Kalb-Ramond Fields}

\author[a,b]{E. Elizalde}
\author[a,b,c]{S.D. Odintsov}
\author[d,e,f]{V.K. Oikonomou}
\author[g]{Tanmoy Paul}

\affiliation[a]{ Institute of Space Sciences (ICE,CSIC) C. Can
Magrans s/n, 08193 Barcelona, Spain}

\affiliation[b]{ Institute of Space Sciences of Catalonia (IEEC),
Barcelona, Spain}

\affiliation[c]{ICREA, Passeig Luis Companys, 23, 08010 Barcelona,
Spain}

\affiliation[d] {Department of Physics, Aristotle University of
Thessaloniki, Thessaloniki 54124, Greece}

\affiliation[e]{Laboratory for Theoretical Cosmology, Tomsk State
University of Control Systems and Radioelectronics, 634050 Tomsk,
Russia (TUSUR)}

\affiliation[f]{Tomsk State Pedagogical University, 634061 Tomsk,
Russia}

\affiliation[g]{Department of Theoretical Physics,\\
Indian Association for the Cultivation of Science,\\
2A $\&$ 2B Raja S.C. Mullick Road,\\
Kolkata - 700 032, India}

\abstract{In this paper we shall study the inflationary aspects of
a logarithmic corrected $R^2$ Starobinsky inflation model, in the
presence of a Kalb-Ramond field in the gravitational action of
$F(R)$ gravity. Our main interest is to pin down the effect of
this rank two antisymmetric tensor field on the inflationary
phenomenology of the $F(R)$ gravity theory at hand. The effects of
the Kalb-Ramond field are expected to be strong during the
inflationary era, however as the Universe expands, the energy
density of the Kalb-Ramond field scales as $\sim a^{-6}$ so dark
matter and radiation dominate over the Kalb-Ramond field effects.
In general, antisymmetric fields constitute the field content of
superstring theories, and thus their effect at the low-energy
limit of the theory is expected to be significant. As we will
show, for a flat Friedmann-Robertson-Walker metric, the
Kalb-Ramond field actually reduces to a scalar field, so it is
feasible to calculate the observational indices of inflation. We
shall calculate the spectral index and the tensor-to-scalar ratio
for the model at hand, by assuming two conditions for the
resulting Kalb-Ramond scalar field, the slow-roll and the
constant-roll condition. As we shall demonstrate, in both the
slow-roll and constant-roll cases, compatibility with the latest
observational data can be achieved. Also the effect of the
Kalb-Ramond field on the inflationary phenomenology is to increase
the amount of the predicted primordial gravitational radiation, in
comparison to the corresponding $f(R)$ gravities, however the
results are still compatible with the observational data.}

\maketitle



\def\pp{{\, \mid \hskip -1.5mm =}}
\def\cL{\mathcal{L}}
\def\be{\begin{equation}}
\def\ee{\end{equation}}
\def\bea{\begin{eqnarray}}
\def\eea{\end{eqnarray}}
\def\tr{\mathrm{tr}\, }
\def\nn{\nonumber \\}
\def\e{\mathrm{e}}

\section{Introduction}

The inflationary paradigm
\cite{Guth:1980zm,Starobinsky:1980te,Linde:1983gd,inflation1,inflation2,inflation4}
is one of the two most successful scenarios that can consistently
describe the primordial era of our Universe, with the second
scenario being the bounce cosmology scenario \cite{bounce}. Both
scenarios can predict a nearly scale invariant power spectrum and
a small amount of gravitational radiation, which are also verified
and tightly constrained by the latest Planck \cite{Akrami:2018odb}
and BICEP2/Keck-Array data \cite{Array:2015xqh}. Due to the
constraints coming from the observations, the number of viable
models of inflation have been narrowed down, and the quest for
modern theoretical cosmologists is to find the an optimal
description for the primordial era than can produce a nearly scale
invariant power spectrum, and a small amount of gravitational
radiation, and at the same time describe with the same model the
late-time acceleration era. Modified gravity
\cite{reviews1,reviews2,reviews3,reviews4,reviews5,reviews6} and
specifically $F(R)$ gravity, serves as a theoretical framework
which can describe successfully both the early and late-time
acceleration eras, see for example the model \cite{Nojiri:2003ft}
for a characteristic model of this sort. In the context of $F(R)$
gravity, it is possible to generate quite successful models of
inflation, with the most well-known being the Starobinsky model
\cite{Starobinsky:1980te}, see also
\cite{Castellanos:2018dub,Wang:2018kly} for a recent modification
of the Starobinsky model. In this line of research, studying
modifications of the standard $R^2$ Starobinsky inflation model,
may provide useful insights with regard to both the early-time and
the late-time acceleration eras. In view of this aspects, in this
work we shall consider a logarithmic-corrected $R^2$ model of the
form,
\begin{eqnarray}
 F(R) = R + \alpha R^2 + \beta R^2\ln{(\beta R)}\, ,
 \label{intro 1}
\end{eqnarray}
with $\alpha$, $\beta$ being constant parameters of the model, in
the presence of a Kalb-Ramond (KR) field $B_{\mu\nu}$ in the
gravitational action of the vacuum $F(R)$ gravity. The logarithmic
$F(R)$ gravity model of Eq. (\ref{intro 1}) is known to provide a
viable early-time phenomenology and a qualitatively consistent
late-time phenomenology \cite{Odintsov:2017hbk}. Actually, due to
fact that logarithmic corrections are induced by one-loop effects
in quantum gravity, there is still much interest to inflationary
phenomenology in the context of logarithmic corrected modified
gravity, see for example
\cite{Odintsov:2018qug,Nojiri:2003ni,Cognola:2005de,Elizalde:2017mrn,Myrzakulov:2014hca,Odintsov:2017hbk,Liu:2018hno}.
So in this work we question the viability of the model in the
presence of this string theory inspired KR field. The presence of
this string-inspired term is motivated by the fact that during the
primordial epoch, quantum gravity or string theory effects may
have a significant imprint on the evolution of the Universe, so in
our case we quantify the quantum epoch's imprint on the evolution
of the Universe, by using this rank two KR antisymmetric tensor
field. In general, antisymmetric tensor fields or equivalently
$p$-forms, constitute the field content of all superstring models,
and in effect these can actually have a realistic impact in the
low-energy limit of the theory \cite{Buchbinder:2008jf}. In
support of this, the Calabi-Yau compactifications of ten
dimensional superstring theories to four dimensions, lead to
low-energy supergravity theories which contain electric and
magnetic fluxes of various $p$-form antisymmetric fields. These
theories are intriguing due to the resolution of the vacuum
degeneracy that the resulting scalar potentials provide. In fact,
the magnetic charges of the low-energy supergravity theory
generate mass terms for the $p$-forms, so in effect one obtains
supersymmetric models which contain antisymmetric tensor fields
\cite{Buchbinder:2008jf}. In this work we shall consider the
effect of a massless antisymmetric KR field on $F(R)$ gravity
inflation, and the effects of a massive antisymmetric field will
be studied in a future work. For a relevant work on the effect of
KR fields in $F(R)$ gravity, see \cite{Das:2018jey}.

As the Universe evolves and cools down, the contribution of the KR
field on the evolutionary process reduces significantly, and at
present day it does not affect the Universe's evolution at all.
Therefore, in this work we shall consider the effects of this
string inspired KR field on the observational indices of
inflation, and specifically on the spectral index of the
primordial curvature perturbations and on the tensor-to-scalar
ratio. On the other hand, the motivation for using a logarithmic
corrected $R^2$ Starobinsky inflation, comes from studies which
include one-loop corrections in higher derivative quantum gravity
\cite{Myrzakulov:2014hca}. Thus the main focus in this work is to
confront phenomenologically the logarithmic-corrected $R^2$
Starobinsky inflation model in the presence of a KR field, with
the observational data coming from the Planck
\cite{Akrami:2018odb} and the BICEP2/Keck-Array collaborations
\cite{Array:2015xqh}. For the calculation, we shall consider the
slow-roll and constant-roll
\cite{Inoue:2001zt,Tsamis:2003px,Kinney:2005vj,Tzirakis:2007bf,
Namjoo:2012aa,Martin:2012pe,Motohashi:2014ppa,Cai:2016ngx,
Motohashi:2017aob,Hirano:2016gmv,Anguelova:2015dgt,Cook:2015hma,
Kumar:2015mfa,Odintsov:2017yud,Odintsov:2017qpp,Lin:2015fqa,Gao:2017uja,Nojiri:2017qvx,Motohashi:2017vdc,Gao:2017owg,Cicciarella:2017nls,Awad:2017ign,Anguelova:2017djf,Ito:2017bnn,Karam:2017rpw,Yi:2017mxs,Mohammadi:2018oku,Gao:2018tdb,Gao:2018cpp,Morse:2018kda,Mohammadi:2018zkf,Boisseau:2018rgy}
evolution cases. As we shall demonstrate, in both cases the
resulting inflationary phenomenology can be compatible with the
observational data, by appropriately choosing the free parameters
of the gravitational model at hand.

The paper is organized as follows: In section II, we briefly
discuss how a general $F(R)$ gravity model can be recast into
Einstein gravity plus a scalar field. In section III we introduce
the KR $F(R)$ gravity model, present the cosmological equations of
motion for a flat Friedmann-Robertson-Walker (FRW) metric and we
present the form of the gravitational equations under the
assumptions of having a constant-roll and a slow-roll condition
imposed on the evolution of the KR field. In section IV the
phenomenological aspects of the model are studied in detail, and
finally the conclusions follow at the end of the paper.

\section{$F(R)$ Gravity in the Einstein Frame}

In this section, we briefly discuss how the higher curvature
$F(R)$ gravity model in four dimensions can be recast into an
Einstein gravity theory with a scalar field. The Jordan frame
vacuum $F(R)$ action has the following form,
\begin{eqnarray}
 S = \int d^4x \sqrt{-g}\Big{(}\frac{F(R)}{2\kappa^2}\Big{)}\, ,
 \label{transformation1}
\end{eqnarray}
where $g$ is the determinant of the metric $g_{\mu\nu}$, $R$
denotes the Ricci scalar and finally $\frac{1}{2\kappa^2} = M_p^2$
where $M_p$ is the Planck mass. Introducing an auxiliary field
$A$, the gravitational action (\ref{transformation1}), the latter
can equivalently be cast as follows,
\begin{eqnarray}
 S = \int d^4x \sqrt{-g}\frac{1}{2\kappa^2}\Big{(}F'(A)(R-A) +
 F(A)\Big{)}\, ,
 \label{transformation2}
\end{eqnarray}
Upon varying the gravitational action with respect to the
auxiliary field $A$, one easily obtains the solution $A = R$.
Plugging back this solution $A=R$ into the action
(\ref{transformation2}), the original action of Eq.
(\ref{transformation1}) can be reproduced. At this stage, we
perform the following conformal transformation of the metric
($g_{\mu\nu}(x)$),
\begin{equation}
 g_{\mu\nu}(x) \longrightarrow \widetilde{g}_{\mu\nu}(x) = e^{-\sqrt{\frac{2}{3}}\kappa\xi} g_{\mu \nu}(x)
 \label{transformation3}
\end{equation}
where $\xi(x)$ is the conformal factor and it is related to the
auxiliary field in the following way $F'(A) =
e^{-[\sqrt{\frac{2}{3}}\kappa\xi]}$. If $R$ and $\tilde{R}$ are
the Ricci scalars corresponding to the metrics $g_{\mu\nu}$ and
$\tilde{g}_{\mu\nu}$ respectively, then these are related as
follows,
\begin{eqnarray}
 R =e^{-\Big{(}\sqrt{\frac{2}{3}}\kappa\xi\Big{)}}\Big{(}\tilde{R} - \kappa^2\tilde{g}^{\mu\nu}\partial_{\mu}\xi\partial_{\nu}\xi
 + 2\kappa\sqrt{\frac{3}{2}}\tilde{\Box}\xi \Big{)}\, .
 \nonumber
\end{eqnarray}
where $\tilde{\Box}$ represents the d' Alembertian operator formed
by the metric tensor $\tilde{g}_{\mu\nu}$. Due to the above
relation between $R$ and $\tilde{R}$, the action
(\ref{transformation2}) can be written as follows,
\begin{eqnarray}
 S = \int d^4x \sqrt{-\tilde{g}}&\bigg[&\frac{1}{2\kappa^2} e^{[\sqrt{\frac{2}{3}}\kappa\xi]}F'(A)\bigg(\tilde{R}
 - \kappa^2\tilde{g}^{\mu\nu}\partial_{\mu}\xi\partial_{\nu}\xi
 + 2\kappa \sqrt{\frac{3}{2}}\tilde{\Box}\xi\bigg)\nonumber\\
 &-&\frac{1}{2\kappa^2}e^{2[\sqrt{\frac{2}{3}}\kappa\xi]}\bigg(AF'(A) - F(A)\bigg)\bigg]
 \label{transformation intermediate}
\end{eqnarray}
Considering $F'(A) > 0$ and using the aforementioned relation
between $\xi$ and $F'(A)$, one obtains the following scalar-tensor
action,
\begin{eqnarray}
 S&=&\int d^4x \sqrt{-\tilde{g}}\bigg[\frac{\widetilde{R}}{2\kappa^2} - \frac{1}{2}\tilde{g}^{\mu\nu}\partial_{\mu}\xi \partial_{\nu}\xi
 - \bigg(\frac{AF'(A) - F(A)}{2\kappa^2F'(A)^2}\bigg)\bigg]
 \label{transformation4}
\end{eqnarray}
Noticeably the field $\xi(x)$ acts as a scalar field with a
potential $V(A(\xi))=\frac{AF'(A) - F(A)}{2\kappa^2F'(A)^{2}}$.
Thus the higher curvature degree of freedom in the Jordan frame,
after a conformal transformation manifests itself in terms of the
scalar field degree of freedom $\xi(x)$ with a potential $V(\xi)$,
which in turn depends on the functional form of $F(R)$. Further it
is also important to note that for $F'(R) < 0$, the kinetic term
of the scalar field $\xi$, as well as the Ricci scalar $R$ in the
above action come with wrong sign, which indicates the existence
of a ghost field. Thus to avoid the ghost like structures, the
derivative of the functional form of the $F(R)$ gravity, namely
$F'(R)$ must be greater than zero. Later we shall show that in the
model we shall study, this condition is indeed satisfied.

\section{The Logarithmic $F(R)$ Gravity Model in the Presence of the Kalb-Ramond Field}

In this section we shall present the logarithmic-corrected $F(R)$
gravity model, in the presence of a KR field and we also employ
the formalism for the case of a flat cosmological evolution. The
vacuum gravitational action of the model is,
\begin{eqnarray}
 S&=&\int d^4x \sqrt{-g}\bigg[\frac{F(R)}{2\kappa^2} - \frac{1}{12}H_{\mu\nu\rho}H^{\mu\nu\rho}\bigg]\nonumber\\
 &=&\int d^4x \sqrt{-g}\bigg[\frac{1}{2\kappa^2}\bigg(R + \alpha R^2 + \beta R^2\ln{(\beta R)}\bigg) - \frac{1}{12}H_{\mu\nu\rho}H^{\mu\nu\rho}\bigg]
 \label{action1}
\end{eqnarray}
where $\frac{1}{2\kappa^2}=M_{p}^2$. Moreover $H_{\mu\nu\alpha}$
is the field strength tensor of the KR field, defined by
$H_{\mu\nu\alpha} = \partial_{[\mu}B_{\nu\alpha]}$. The field
strength tensor $H_{\mu\nu\alpha}$ is invariant under the KR gauge
transformation, $B_{\mu\nu}\rightarrow B_{\mu\nu} +
\partial_{\mu}\omega_{\nu}-\partial_{\nu}\omega_{\mu}$ and consequently,  the action (\ref{action1}) is also invariant under this gauge transformation. However the
form of $F(R)$ gravity clearly indicates that in the regime of
$\alpha$, $\beta$ $> 0$, $F'(R)$ comes with positive sign (for
$R>0$) which in turn renders the model ghost-free. Thus in the
rest of this work we shall consider only positive values for the
parameters $\alpha$ and $\beta$. By using the formalism of the
previous section we can rewrite the action (\ref{action1})  into a
scalar-tensor form,
\begin{eqnarray}
 S&=&\int d^4x \sqrt{-\tilde{g}}\bigg[\frac{\widetilde{R}}{2\kappa^2} - \frac{1}{2}\tilde{g}^{\mu\nu}\partial_{\mu}\xi \partial_{\nu}\xi
 - V(\xi)\nonumber\\
 &-&\frac{1}{12}e^{-\sqrt{\frac{2}{3}}\kappa \xi} H_{\mu\nu\rho}H_{\alpha\beta\delta}
 \tilde{g}^{\mu\alpha}\tilde{g}^{\nu\beta}\tilde{g}^{\rho\delta}\bigg]\,
 ,
 \label{action3}
\end{eqnarray}
with the scalar field being related to the higher curvature
degrees of freedom as follows,
\begin{eqnarray}
 \sqrt{\frac{2}{3}}\kappa\xi(R) = \ln{\bigg[\frac{1}{1 + (2\alpha+\beta)R + 2\beta R\ln{(\beta
 R)}}\bigg]}\, ,
 \label{scalar field higher curvature}
\end{eqnarray}
and the scalar potential $V(\xi)$ has the following form,
\begin{eqnarray}
 V(\xi(R)) = \frac{1}{2\kappa^2}\bigg[\frac{\big[(\alpha+\beta)R^2 + \beta R^2\ln{(\beta R)}\big]}
 {\big[1 + (2\alpha+\beta)R + 2\beta R\ln{(\beta
 R)}\big]^2}\bigg]\, ,
 \label{potential}
\end{eqnarray}
and finally with $\xi=\xi(R)$ being given in Eq. (\ref{scalar
field higher curvature}). We need to note that the transformations
of the KR field between Jordan and Einstein frames has been worked
out for general $f(R)$ gravity theories in Ref.
\cite{Das:2018jey}. The potential has a minimum value at,
\begin{eqnarray}
 2\alpha + 3\beta + 2\beta\ln{(\beta <R>_{min})} = 0\, ,
 \label{minima}
\end{eqnarray}
and a maximum value which is,
\begin{eqnarray}
 <R>_{max} = 1/\beta\, .
 \label{maxima}
\end{eqnarray}
Moreover at the large curvature regime $R \gg 1$, $V(R)$ behaves
as $\frac{1}{8\kappa^2\beta\ln{(\beta R)}}$ and it thus slowly
approaches the value zero as $R\rightarrow \infty$. Furthermore,
from Eq. (\ref{potential}), one obtains the first and second
derivative of the potential, which are,
\begin{eqnarray}
 \frac{dV}{d\xi}&=&\frac{dV}{dR}\frac{dR}{d\xi}\nonumber\\
 &=&\frac{\sqrt{\frac{2}{3}}}{2\kappa} \bigg[\frac{(-R + \beta R^2)}{[1 + (2\alpha + \beta)R + 2\beta R\ln{(\beta R)}]^2}\bigg]
 \label{first derivative}
\end{eqnarray}
and
\begin{equation}
 \frac{d^2V}{d\xi^2} =
 \frac{1}{3}\bigg[\frac{-3R - 2\alpha R^2 + \beta R^2 - 2\beta R^2\ln{(\beta R)}}{[1 + (2\alpha + \beta)R + 2\beta R\ln{(\beta R)}]^2}
 + \frac{1}{2\alpha + 3\beta + 2\beta \ln{(\beta R)}}\bigg]\, ,
 \label{second derivative}
\end{equation}
which we will need later on in order to determine the spectral
index and tensor-to-scalar ratio.  However it is evident from Eq.
(\ref{action3}) that due to the presence of the scalar field
$\xi(x)$ (from higher curvature degrees of freedom), the kinetic
term of the KR field becomes non-canonical. In order to make it
canonical, we redefine the field as follows,
\begin{eqnarray}
 B_{\mu\nu} \longrightarrow \widetilde{B}_{\mu\nu} = e^{-\frac{1}{2}\sqrt{\frac{2}{3}}\kappa \xi}
 B_{\mu\nu}\, .
 \label{redefine}
 \end{eqnarray}
In terms of the redefined field, the final form of the
scalar-tensor action is the following,
 \begin{eqnarray}
 S = \int d^4x \sqrt{-\tilde{g}}\bigg[\frac{\widetilde{R}}{2\kappa^2} - \frac{1}{2}\tilde{g}^{\mu\nu}\partial_{\mu}\xi \partial_{\nu}\xi
 - V(\xi) - \frac{1}{12}
 \tilde{H}_{\mu\nu\rho}\tilde{H}^{\mu\nu\rho}\bigg]\, ,
 \label{action4}
\end{eqnarray}
and we shall consider the case where $\kappa\tilde{B}_{\mu\nu}<1$,
and also $V(\xi)$ is given in Eq. (\ref{potential}). Accordingly,
we shall determine the cosmological field equations for the
scalar-tensor model, from which one can go back to the original
$F(R)$ gravity model of Eq. (\ref{action1})  by using an inverse
conformal transformation.

\subsection{Cosmological Field Equations for the Scalar-tensor Model}

In order to obtain the field equations of the scalar-tensor (ST)
action (\ref{action4}), first we determine the energy-momentum
tensor for $\xi(x)$ and $\tilde{B}_{\mu\nu}(x)$, which are,
\begin{eqnarray}
 T_{\mu\nu}[\xi]&=&\frac{2}{\sqrt{-\tilde{g}}}
 \frac{\delta}{\delta \tilde{g}^{\mu\nu}}\bigg[\sqrt{-\tilde{g}}\bigg(\frac{1}{2}\tilde{g}^{\alpha\beta}\partial_{\alpha}\xi\partial_{\beta}\xi + V(\xi)\bigg)\bigg]\nonumber\\
 &=&\partial_{\mu}\xi\partial_{\nu}\xi - \tilde{g}_{\mu\nu}\bigg(\frac{1}{2}\tilde{g}^{\alpha\beta}\partial_{\alpha}\xi\partial_{\beta}\xi + V(\xi)\bigg)
 \label{em tensor1}
\end{eqnarray}
and
\begin{eqnarray}
 T_{\mu\nu}[\tilde{B}]&=&\frac{2}{\sqrt{-\tilde{g}}}
 \frac{\delta}{\delta \tilde{g}^{\mu\nu}}\bigg[\frac{1}{12}\sqrt{-\tilde{g}}\tilde{g}^{\mu\alpha}\tilde{g}^{\nu\beta}\tilde{g}^{\lambda\gamma}
 \tilde{H}_{\mu\nu\lambda}\tilde{H}_{\alpha\beta\gamma}\bigg]\nonumber\\
 &=&\frac{1}{6}\bigg[3\tilde{g}_{\nu\rho}\tilde{H}_{\alpha\beta\mu}\tilde{H}^{\alpha\beta\rho}
 - \frac{1}{2}\tilde{g}_{\mu\nu}\tilde{H}_{\alpha\beta\gamma}\tilde{H}^{\alpha\beta\gamma}\bigg]
 \label{em tensor2}
\end{eqnarray}
respectively. Our motivation is to investigate whether the $F(R)$
model that we chose in the present context can be considered as a
viable inflationary model, in view of the observational
constraints of $Planck$ 2018 and of the BICEP2 Keck-Array data. We
shall consider a flat FRW metric of the form,
\begin{equation}
 \tilde{ds}^2=-dt^2 + a^2(t)\big[dx^2 + dy^2 + dz^2\big]
 \label{4d metric}
\end{equation}
where $t$ and $a(t)$ are cosmic time and scale factor
respectively. However before presenting the field equations, we
want to emphasize that due to antisymmetric nature of the KR
field, the tensor $\tilde{H}_{\mu\nu\lambda}$ has four independent
components in the context of four dimensional spacetime, which can
be expressed as follows,
\begin{eqnarray}
 \tilde{H}_{012} = h_1,\,\,\,\,\tilde{H}^{012} = h^1\, ,\nonumber\\
 \tilde{H}_{013} = h_2,\,\,\,\,\tilde{H}^{013} = h^2\, ,\nonumber\\
 \tilde{H}_{023} = h_3,\,\,\,\,\tilde{H}^{023} = h^3\, ,\nonumber\\
 \tilde{H}_{123} = h_4,\,\,\,\,\tilde{H}^{123} = h^4\, ,
\label{independent}
\end{eqnarray}
At this stage, it is worth mentioning that due to the presence of
the four independent components, the KR field tensor
$\tilde{H}_{\mu\nu\alpha}$ can be equivalently expressed by a
vector field (which has also four independent components in four
dimensions) \cite{buchbinder} as $\tilde{H}_{\mu\nu\alpha} =
\varepsilon_{\mu\nu\alpha\beta}\Upsilon^{\beta}$, with
$\Upsilon^{\beta}$ being the vector field.  Therefore, Eq.
(\ref{independent}) for the FRW metric can yield the components of
the energy momentum tensors $T_{\mu\nu}[\xi]$ and
$T_{\mu\nu}[\tilde{B}]$, which are,
\begin{eqnarray}
 T_{00}[\xi]&=&\frac{1}{2}\dot{\xi}^2 + V(\xi)\nonumber\\
 T_{11}[\xi]&=&T_{22}[\xi] = T_{33}[\xi] = a(t)^2\bigg[\frac{1}{2}\dot{\xi}^2 - V(\xi)\bigg]
 \nonumber
\end{eqnarray}
and
\begin{eqnarray}
 T_{00}[\tilde{B}]&=&\frac{1}{2}\bigg[-h_1h^1 - h_2h^2 - h_3h^3 + h_4h^4\bigg]\nonumber\\
 T_{11}[\tilde{B}]&=&\frac{1}{2}a(t)^2\bigg[h_1h^1 + h_2h^2 - h_3h^3 + h_4h^4\bigg]\nonumber\\
 T_{22}[\tilde{B}]&=&\frac{1}{2}a(t)^2\bigg[h_1h^1 - h_2h^2 + h_3h^3 + h_4h^4\bigg]\nonumber\\
 T_{33}[\tilde{B}]&=&\frac{1}{2}a(t)^2\bigg[-h_1h^1 + h_2h^2 + h_3h^3 + h_4h^4\bigg]
 \nonumber
\end{eqnarray}
\begin{eqnarray}
 T_{10}[\tilde{B}]&=&-h_4h^3,\,\,\,\,T_{20}[\tilde{B}] = h_4h^2\nonumber\\
 T_{30}[\tilde{B}]&=&-h_4h^1,\,\,\,\,T_{12}[\tilde{B}] = a(t)^2h_2h^3\nonumber\\
 T_{13}[\tilde{B}]&=&-a(t)^2h_1h^3,\,\,\,\,T_{23}[\tilde{B}] = a(t)^2h_1h^2
 \nonumber
\end{eqnarray}
where the fields are taken to be homogeneous in space. In effect,
the off-diagonal Friedmann equations read,
\begin{eqnarray}
 h_4h^3 = h_4h^2 = h_4h^1 = h_2h^3 = h_1h^3 = h_1h^2 = 0\, ,
 \label{off einstein equation}
\end{eqnarray}
which have the following solution,
\begin{eqnarray}
 h_1 = h_2 = h_3 = 0,\,\,\,\, h_4 \neq 0\, .
 \label{sol off einstein equation}
\end{eqnarray}
Using this solution, one easily obtains the total energy density
and pressure for the matter fields ($\xi$, $\tilde{B}_{\mu\nu}$)
which are,
\begin{align}\label{newequations3a}
& \rho_T = \bigg[\frac{1}{2}\dot{\xi}^2 + V(\xi) +
\frac{1}{2}h_4h^4\bigg]\, ,\\ \notag & p_T =
\bigg[\frac{1}{2}\dot{\xi}^2 - V(\xi) + \frac{1}{2}h_4h^4\bigg]\,
,
\end{align}
with the ``dot'' denoting differentiation with respect to the
cosmic time. In effect, the diagonal Friedmann equations turn out
to be,
\begin{eqnarray}
 H^2 = \frac{\kappa^2}{3}\bigg[\frac{1}{2}\dot{\xi}^2 + V(\xi) +
 \frac{1}{2}h_4h^4\bigg]\, ,
 \label{einstein equation1}
\end{eqnarray}
and
\begin{eqnarray}
 2\dot{H} + 3H^2 + \kappa^2\bigg[\frac{1}{2}\dot{\xi}^2 - V(\xi) + \frac{1}{2}h_4h^4\bigg] =
 0\, ,
 \label{einstein equation2}
\end{eqnarray}
where $H=\frac{\dot{a}}{a}$ denotes the Hubble parameter of the
scalar-tensor model. Furthermore, the field equations for the KR
field ($\tilde{B}_{\mu\nu}$) and the scalar field ($\xi$) are
given by,
\begin{eqnarray}
 \tilde{\nabla}_{\mu}\tilde{H}^{\mu\nu\lambda} = \frac{1}{a^3(t)}\partial_{\mu}\bigg[a^3(t)\tilde{H}^{\mu\nu\lambda}\bigg] =
 0\, ,
 \label{KR equation}
\end{eqnarray}
and
\begin{eqnarray}
 \ddot{\xi} + 3H\dot{\xi} + \frac{\partial V}{\partial\xi}  = 0\,
 ,
 \label{scalar equation}
\end{eqnarray}
respectively. However, first we investigate Eq. (\ref{KR
equation}). As we will see shortly, the only information that can
be obtained from Eq. (\ref{KR equation}) is that the non-zero
component of $\tilde{H}_{\mu\nu\alpha}$ (i.e
$\tilde{H}_{123}=h_4$) depends solely on the cosmic time $t$. This
can easily by seen,  by expanding Eq. (\ref{KR equation}) as
follows,
 \begin{eqnarray}
 & \partial_{0}&\bigg[a^3(t)\tilde{H}^{0\nu\lambda}\bigg] + \partial_{1}\bigg[a^3(t)\tilde{H}^{1\nu\lambda}\bigg]\nonumber\\
 &+\partial_{2}&\bigg[a^3(t)\tilde{H}^{2\nu\lambda}\bigg] + \partial_{3}\bigg[a^3(t)\tilde{H}^{3\nu\lambda}\bigg] =
 0\, .
 \label{app2 2}
\end{eqnarray}
 Therefore for,
\begin{itemize}
 \item $\nu = 2$ and $\lambda = 3$, Eq. (\ref{app2 2}) becomes
 \begin{eqnarray}
  &\partial_{t}&\bigg[a^3(t)\tilde{H}^{023}\bigg] + \partial_{x}\bigg[a^3(t)\tilde{H}^{123}\bigg]\nonumber\\
 &+\partial_{y}&\bigg[a^3(t)\tilde{H}^{223}\bigg] + \partial_{z}\bigg[a^3(t)\tilde{H}^{323}\bigg] = 0
 \label{app2 3}
 \end{eqnarray}
Due to the antisymmetric nature of the KR field, the last two
terms of the above equation vanish identically. Furthermore, from
Eq. (\ref{sol off einstein equation}), we get $\tilde{H}^{023} =
0$. As a result, only the second term of Eq. (\ref{app2 3})
survives, from which we may conclude that the non-zero component
of the KR field ($\tilde{H}^{123}$) is independent of the spatial
coordinates $x$ i.e. $\partial_{x}\bigg[\tilde{H}^{123}\bigg] =
0$.

\item For $\nu = 1$ and $\lambda = 3$, Eq. (\ref{app2 2}) becomes,
\begin{eqnarray}
  &\partial_{t}&\bigg[a^3(t)\tilde{H}^{013}\bigg] + \partial_{x}\bigg[a^3(t)\tilde{H}^{113}\bigg]\nonumber\\
 &+\partial_{y}&\bigg[a^3(t)\tilde{H}^{213}\bigg] + \partial_{z}\bigg[a^3(t)\tilde{H}^{313}\bigg] = 0
 \label{app2 4}
 \end{eqnarray}
 Here the third term survives, which ensures that $\tilde{H}^{123}$ is independent of $y$.

 \item For $\nu = 1$ and $\lambda = 2$, Eq. (\ref{app2 2}) becomes,
 \begin{eqnarray}
  &\partial_{t}&\bigg[a^3(t)\tilde{H}^{012}\bigg] + \partial_{x}\bigg[a^3(t)\tilde{H}^{112}\bigg]\nonumber\\
 &+\partial_{y}&\bigg[a^3(t)\tilde{H}^{212}\bigg] + \partial_{z}\bigg[a^3(t)\tilde{H}^{312}\bigg] = 0
 \label{app2 5}
 \end{eqnarray}
 where the fourth term gives $\partial_{z}\bigg[\tilde{H}^{123}\bigg] = 0$.

\end{itemize}
Therefore, it is clear that the non-zero component of the KR field
i.e $\tilde{H}^{123}$ depends solely on the cosmic time
coordinate, which is also expected from the gravitational field
equations. In view of these results, now we turn our focus to the
other field equations. Differentiating both sides (with respect to
$t$) of Eq. (\ref{einstein equation1}), we get,
\begin{eqnarray}
6H\dot{H} = \kappa^2\bigg[\dot{\xi}\ddot{\xi} + V'(\xi)\dot{\xi} +
\frac{1}{2}\frac{d}{dt}(h_4h^4)\bigg] \nonumber
\end{eqnarray}
Furthermore,  Eqs. (\ref{einstein equation1}) and (\ref{einstein
equation2}) lead to the expression as $2\dot{H} =
-\kappa^2\big[\dot{\xi}^2 + h_4h^4\big]$. Plugging back this
expression of $\dot{H}$ into the above equation and using the
scalar field equation of motion, we obtain the following cosmic
evolution of $h_4h^4$,
\begin{eqnarray}
 \frac{d}{dt}(h_4h^4) = -6H h_4h^4
 \label{evolution of KR energy density}
\end{eqnarray}
Recall that the term $\frac{1}{2}h_4h^4$ represents the energy
density $\tilde{\rho}_{KR}$ corresponding to the KR field.
Therefore Eq. (\ref{evolution of KR energy density}) can be
alternatively written as follows,
\begin{equation}\label{newequation34a}
\frac{d\tilde{\rho}_{KR}}{dt} = -6H\tilde{\rho}_{KR}\, ,
\end{equation}
\begin{eqnarray}
 \tilde{\rho}_{KR} = \frac{h_0}{2a^6}\, ,
 \label{solution of KR energy density}
\end{eqnarray}
with $h_0$ being an integration constant which must take only
positive values in order to get a real valued solution for
$h^4(t)$. In addition, Eq. (\ref{solution of KR energy density})
clearly indicates that the energy density of the KR field
($\tilde{\rho}_{KR}$) is proportional to $1/a^6$ and in effect,
$\tilde{\rho}_{KR}$ decreases as the Universe expands, in a faster
rate in comparison to matter ($\propto 1/a^3$) and radiation
($\propto 1/a^4$) energy density respectively. With the solution
of $h_4h^4$ at hand (in terms of scale factor, see Eq.
(\ref{solution of KR energy density})), there remain two
independent equations which are the following,
\begin{eqnarray}
 H^2 = \frac{\kappa^2}{3}\bigg[\frac{1}{2}\dot{\xi}^2 + V(\xi) +
 \frac{1}{2}\frac{h_0}{a^6}\bigg]\, ,
 \label{independent equation1}
\end{eqnarray}
and
\begin{eqnarray}
 \ddot{\xi} + 3H\dot{\xi} +  \frac{\partial V}{\partial\xi} = 0\,
 .
 \label{independent equation2}
 \end{eqnarray}
Here, we need to mention that Eqs. (\ref{independent equation1}),
(\ref{independent equation2}) match with the field equations when
$\tilde{H}_{\mu\nu\alpha}$ is expressed by using it's vector
representation i.e.
$\tilde{H}_{\mu\nu\alpha}=\varepsilon_{\mu\nu\alpha\beta}\Upsilon^{\beta}$.
In the Appendix we provide the details of this equivalence between
the two field equations. This confirms the equivalence between the
two representations (when $\tilde{H}_{\mu\nu\alpha}$ is not
represented by a vector field) at the level of equations of
motion, which is also in agreement with Ref. \cite{buchbinder}.
Since we are interested in finding the observational indices of
inflation, and specifically the spectral index and the
tensor-to-scalar ratio, we shall make some simplifications on the
evolution of the KR, field, and specifically we shall assume that
it obeys the slow-roll or the constant-roll condition.

Let us first consider the slow-roll case, in which case the
assumption which is made is the following,
 \begin{eqnarray}
 V(\xi) \gg \frac{1}{2}\dot{\xi}^2
 \label{slow roll approximation}
\end{eqnarray}
Under these conditions, the field equations become,
\begin{eqnarray}
 H^2 = \frac{\kappa^2}{3}\bigg[V(\xi) + \frac{h_0}{2a^6}\bigg]~~~~,
 \label{slow roll equation1a}
\end{eqnarray}
\begin{eqnarray}
 2\dot{H} + 3H^2 + \kappa^2\big[-V(\xi) + \frac{h_0}{2a^6}\big]\,
 ,
 \label{slow roll equation intermediate}
\end{eqnarray}
and
\begin{eqnarray}
 3H\dot{\xi} +  \frac{\partial V}{\partial\xi} = 0\, .
 \label{slow roll equation2a}
\end{eqnarray}
Therefore from the Einstein's equations, one obtains $\dot{H} =
-\frac{\kappa^2h_0}{2a^6}$, which will be used later on. On the
other hand, the constant-roll condition is materialized by the
following condition,
 \begin{eqnarray}
  \frac{\ddot{\xi}}{H\dot{\xi}} = \gamma\, ,
  \label{constant roll approximation}
 \end{eqnarray}
where $\gamma$ is a real parameter. Such a framework interpolates
between the slow-roll inflation, for which $\ddot{\xi} \simeq 0$
and the ultra-slow-roll inflation, satisfying $V'(\xi) = 0$ over a
range of field values. These two regimes are respectively
reproduced by $\gamma \simeq 0$ and $\gamma = -3$. However
integrating Eq. (\ref{constant roll approximation}) yields the
following expression for the velocity of the scalar field (in
terms of the scale factor)  $\dot{\xi} = Aa^{\gamma}$ ($A$ be the
integration constant) \cite{Martin:2012pe}. With these expressions
of $\dot{\xi}$ and $\ddot{\xi}$, the field equations take the
following form,
\begin{eqnarray}
 H^2 = \frac{\kappa^2}{3}\bigg[\frac{A^2}{2}a^{2\gamma} + V(\xi) + \frac{h_0}{2a^6}\bigg]\, ,
 \label{constant roll equation1a}
\end{eqnarray}
\begin{eqnarray}
 2\dot{H} + 3H^2 + \kappa^2\bigg[\frac{A^2}{2}a^{2\gamma} - V(\xi) +
 \frac{h_0}{2a^6}\bigg]\, ,
 \label{constant roll equation intermediate}
\end{eqnarray}
and
\begin{eqnarray}
 \big(\gamma + 3\big) H\dot{\xi} +  \frac{\partial V}{\partial\xi} = 0\, .
 \label{constant roll equation2a}
\end{eqnarray}
It may be observed that when $\gamma$ and $A$ tend to zero, the
constant-roll field equations match with those corresponding to
the slow-roll approximation.  However using Eqs. (\ref{constant
roll equation1a}) and (\ref{constant roll equation intermediate}),
we get $\dot{H} = -\frac{\kappa^2A^2}{2}a^{2\gamma} -
\frac{\kappa^2h_0}{2a^6}$, thereby it is clear that the presence
of the constant $A$ changes the rate of Hubble parameter in
comparison to the slow-roll approximation case, where $\dot{H} =
-\frac{\kappa^2h_0}{2a^6}$. Now a major question to address is at
the level of a background FRW evolution, and in the context of a
linear perturbation theory, how is the massless KR field
distinguished from the massless scalar field. We address this
issue in detail in the following subsection.

\subsection{Perturbation Equations of the KR Theory}

The second rank antisymmetric Kalb-Ramond field in $D$-spacetime
dimensions has a total of $D(D-1)/2$ independent components, but
due to the fact that the time derivatives of only the spatial
components of the Kalb-Ramond field appear in the Lagrangian, and
in addition the transformation $B_{\mu\nu} \rightarrow B_{\mu\nu}
+
\partial_{[\mu}\omega_{\nu]}$ leaves the KR field Lagrangian
invariant and in conjunction with the fact that a change of the
gauge field by $\omega_{\mu} \rightarrow \omega_{\mu} +
\partial_{\mu}\varphi$ keeps the gauge transformation unchanged,
the actual number of degrees of freedom of a $D$-dimensional KR
field is given by $\frac{(D-1)(D-4)}{2} + 1$
\cite{Chakraborty:2017uku}.

Thereby in four dimensions, the KR field has a single degree of
freedom which enables one to write down the KR field tensor
$H_{\mu\nu\rho}$ in terms of a massless scalar field ($Z$) such
that, $H_{\mu\nu\rho} =
\epsilon_{\mu\nu\rho\sigma}\partial^{\sigma}Z$,
 with $\epsilon_{\mu\nu\rho\sigma} = \sqrt{-g}[\mu\nu\rho\sigma]$.\\
Now we are going to investigate whether the perturbed up to first
order cosmological field equations are equivalent when
$H_{\mu\nu\rho}$ is represented or not represented by a massless
scalar field. This investigation is important as we will use this
equivalence in order to determine the slow-roll indices in the
next section. Recall that, the background spacetime is described
by a spatially homogeneous and isotropic FRW metric given in Eq.
(\ref{4d metric}). For this metric, the background field
equations, when the KR field tensor is not represented by a
massless scalar field, are given in Eqs. (\ref{einstein
equation1})-(\ref{scalar equation}).

\subsubsection{The Case of a Massless Scalar Field}

The action with a massless scalar field denoted as $Z$, instead of
the Kalb-Ramond field, is given by,
\begin{eqnarray}
 S = \int d^4x \sqrt{-\tilde{g}} \bigg[\frac{\tilde{R}}{2\kappa^2} - \frac{1}{2}\tilde{g}^{\mu\nu}\partial_{\mu}\xi\partial_{\nu}\xi - V(\xi)
 - \frac{1}{2}\tilde{g}^{\mu\nu}\partial_{\mu}Z\partial_{\nu}Z\bigg]
 \label{ref_action1}
\end{eqnarray}
where $\xi$ is the scalar field which arises from higher curvature
degrees of freedom. The action leads to the following background
Friedmann equations,
\begin{eqnarray}
 H^2 = \frac{\kappa^2}{3} \bigg[\frac{1}{2}\dot{\xi}^2 + V(\xi) + \frac{1}{2}\dot{Z}^2\bigg]\nonumber\\
 2\dot{H} + 3H^2 + \kappa^2\bigg[\frac{1}{2}\dot{\xi}^2 + V(\xi) + \frac{1}{2}\dot{Z}^2\bigg]
 \label{ref_einstein_eqn1}
\end{eqnarray}
where we made the assumption that $Z$ is spatially homogeneous.
Moreover the field equations for $\xi(t)$ and $Z(t)$ are,
\begin{eqnarray}
 \ddot{\xi} + 3H\dot{\xi} + \frac{\partial V}{\partial \xi} = 0
 \label{ref_scalar_eqn1}
\end{eqnarray}
and
\begin{eqnarray}
 \ddot{Z} + 3H\dot{Z} = 0
 \label{ref_second_scalar_eqn1}
\end{eqnarray}
respectively. It can be shown that the background field equations
(\ref{ref_einstein_eqn1}-\ref{ref_second_scalar_eqn1}) match with
the field equations (\ref{einstein equation1})-(\ref{scalar
equation}) by making use of the relation $H_{\mu\nu\rho} =
\epsilon_{\mu\nu\rho\sigma}\partial^{\sigma}Z$.

The demonstration goes as follows: the aforementioned relation
between $H_{\mu\nu\rho}$ and $Z$ immediately leads to the solution
$h_1 = h_2 = h_3 = 0$ due to the fact that $Z$ is spatially
homogeneous. With these solutions, one easily obtains $h_4h^4 =
\dot{Z}^2$ from which the equivalence between the gravitational
field equations (\ref{ref_einstein_eqn1}) and (\ref{einstein
equation1})-(\ref{scalar equation}) is established. Finally the
expression $h_4h^4 = \dot{Z}^2$ also validates the equivalence
between the conservation equations for $Z(t)$ and for the KR field
respectively.

\subsubsection{First Order Perturbed Friedmann Equations}

We now consider a perturbed FRW spacetime containing scalar and
tensor type perturbations \cite{Hwang:2005hb}:
\begin{eqnarray}
 d\tilde{s}^2 = -(1+2\alpha)dt^2 - 2a\beta_{,i}dtdx^{i} + a^2 \bigg[\delta_{ij} + 2\Psi \delta_{ij} + 2\gamma_{,i,j} + 2C_{ij}\bigg]dx^idx^j
 \label{ref_metric2}
\end{eqnarray}
where $i,j$ denote the spatial indices, $\delta_{ij}$ is the
Kronecker symbol and set the background spatial metric. The
spacetime dependent variables $\alpha$, $\beta$, $\gamma$, $\Psi$
are the scalar type perturbations while $C_{\alpha\beta}$ is the
tensorial one.

\subsubsection{With Kalb-Ramond field}

In this subsection, we will determine the first order perturbed
equations in the presence of a second rank antisymmetric
Kalb-Ramond field with the action given by : $S = \int
d^4x\sqrt{-\tilde{g}}\big[\frac{\tilde{R}}{2\kappa^2} -
\frac{1}{2}\tilde{g}^{\mu\nu}\partial_{\mu}\xi\partial_{\nu}\xi -
V(\xi) -
\frac{1}{12}\tilde{H}_{\mu\nu\rho}\tilde{H}^{\mu\nu\rho}\big]$.
However, in view of the perturbations (as shown in
Eq.(\ref{ref_metric2})), the perturbed components of the
energy-momentum tensor are given by the following expressions :
\begin{eqnarray}
 \delta T_{0}^{0}&=&-\bigg[\frac{1}{2}h_4\delta h^4 + \frac{1}{2}h^4\delta h_4 + \dot{\xi}\delta\dot{\xi} + V'(\xi)\delta \xi\bigg]\nonumber\\
 \delta T_{i}^{0}&=&-\bigg[h_4h^4 + \dot{\xi}^2\bigg]v_{,i}\nonumber\\
 \delta T_{i}^{j}&=&\delta^{i}_{j} \bigg[\frac{1}{2}h_4\delta h^4 + \frac{1}{2}h^4\delta h_4 + \dot{\xi}\delta\dot{\xi} + V'(\xi)\delta \xi\bigg]
 \label{ref_perturbed_em1}
\end{eqnarray}
where $\delta h_4$ and $\delta \xi$ are the perturbations of KR
field and of the scalar field $\xi$ respectively. The above
expressions of $\delta T^{\mu}_{\nu}$ along with the introduction
$k = 3H\alpha - 3\dot{\Psi} - \frac{1}{a^2}\nabla^2\chi$ (where
$\nabla$ denotes the Laplacian operator with respect to the
background metric) yield the first order perturbed gravitational
equations as follows \cite{Hwang:2005hb},
\begin{eqnarray}
 \frac{1}{a^2} \nabla^2\Psi + Hk = -\frac{\kappa^2}{2} \bigg[\frac{1}{2}h_4\delta h^4 + \frac{1}{2}h^4\delta h_4 + \dot{\xi}\delta\dot{\xi} + V'(\xi)\delta \xi\bigg]
 \label{ref_perturbed_eqn1}
\end{eqnarray}
\begin{eqnarray}
 k + \frac{1}{a^2}\nabla^2\chi = \frac{3}{2}\kappa^2 a \big[h_4h^4 + \dot{\xi}^2\big]v
 \label{ref_perturbed_eqn2}
\end{eqnarray}
\begin{eqnarray}
 \dot{\chi} + H\chi - \alpha - \Psi = 0
 \label{ref_perturbed_eqn3}
\end{eqnarray}
\begin{eqnarray}
 \dot{k} + 2Hk + \bigg(3\dot{H} + \frac{1}{a^2}\nabla^2\bigg)\alpha = \frac{\kappa^2}{2}
 \bigg[2h_4\delta h^4 + 2h^4\delta h_4 + 4\dot{\xi}\delta\dot{\xi} - 2V'(\xi)\delta \xi\bigg]
 \label{ref_perturbed_eqn4}
\end{eqnarray}
\begin{eqnarray}
 \ddot{C}^{i}_{j} + 3H\dot{C}^{i}_{j} - \frac{1}{a^2}\nabla^2C^{i}_{j} = 0
 \label{ref_perturbed_eqn5}
\end{eqnarray}
where $\chi = a[\beta + a\dot{\gamma}]$. Moreover the perturbed
equations for $\xi$ and for the KR field are given by,
\begin{eqnarray}
 \delta\ddot{\xi} + 3H\delta\dot{\xi} - \frac{1}{a^2}\nabla^2(\delta \xi) + V''(\xi)\delta \xi = \dot{\xi}[k+\dot{\alpha}] + [\ddot{\xi} - V'(\xi)]\alpha
 \label{ref_perturbed_eqn6}
\end{eqnarray}
and
\begin{eqnarray}
 \frac{1}{2}\big[\dot{h}_4\delta\dot{h}^4 + \dot{h}^4\delta\dot{h}_4\big] + 3H\big[h_4\delta h^4 + h^4\delta h_4\big] - h_4h^4\big[k + \dot{\alpha} - 3H\alpha\big] = 0
 \label{ref_perturbed_eqn7}
\end{eqnarray}
respectively, where we consider $\delta h_1=\delta h_2=\delta h_3=\delta h_4=0$.\\
Having presented the perturbed equations in the presence of the KR
field, now we turn our focus to the case of a massless scalar
field.

\subsubsection{With massless scalar field}

In this case, the action is given by Eq. (\ref{ref_action1}). For
such an action, the perturbed components of the energy-momentum
tensor take the following form,
\begin{eqnarray}
 \delta T_{0}^{0}&=&-\bigg[\dot{Z}\delta\dot{Z} + \dot{\xi}\delta\dot{\xi} + V'(\xi)\delta \xi\bigg]\nonumber\\
 \delta T_{i}^{0}&=&-\bigg[\dot{Z}^2 + \dot{\xi}^2\bigg]v_{,i}\nonumber\\
 \delta T_{i}^{j}&=&\delta^{i}_{j} \bigg[\dot{Z}\delta\dot{Z} + \dot{\xi}\delta\dot{\xi} + V'(\xi)\delta \xi\bigg]
 \label{ref_perturbed_em2}
\end{eqnarray}
where $\delta Z$ denotes the perturbation of the massless scalar
field $Z$. Combining the above expressions with the previous
definitions of $k$ and $\chi$, we get the perturbed gravitational
equations,
\begin{eqnarray}
 \frac{1}{a^2} \nabla^2\Psi + Hk = -\frac{\kappa^2}{2} \bigg[\dot{Z}\delta\dot{Z} + \dot{\xi}\delta\dot{\xi} + V'(\xi)\delta \xi\bigg]
 \label{ref_perturbed_eqn8}
\end{eqnarray}
\begin{eqnarray}
 k + \frac{1}{a^2}\nabla^2\chi = \frac{3}{2}\kappa^2 a \big[\dot{Z}^2 + \dot{\xi}^2\big]v
 \label{ref_perturbed_eqn9}
\end{eqnarray}
\begin{eqnarray}
 \dot{\chi} + H\chi - \alpha - \Psi = 0
 \label{ref_perturbed_eqn10}
\end{eqnarray}
\begin{eqnarray}
 \dot{k} + 2Hk + \bigg(3\dot{H} + \frac{1}{a^2}\nabla^2\bigg)\alpha = \kappa^2
 \bigg[2\dot{Z}\delta\dot{Z} + 2\dot{\xi}\delta\dot{\xi} - V'(\xi)\delta \xi\bigg]
 \label{ref_perturbed_eqn11}
\end{eqnarray}
\begin{eqnarray}
 \ddot{C}^{i}_{j} + 3H\dot{C}^{i}_{j} - \frac{1}{a^2}\nabla^2C^{i}_{j} = 0
 \label{ref_perturbed_eqn12}
\end{eqnarray}
Moreover the perturbed equations for $\xi$ and for $Z$ are given
by,
\begin{eqnarray}
 \delta\ddot{\xi} + 3H\delta\dot{\xi} - \frac{1}{a^2}\nabla^2(\delta \xi) + V''(\xi)\delta \xi = \dot{\xi}[k+\dot{\alpha}] + [\ddot{\xi} - V'(\xi)]\alpha
 \label{ref_perturbed_eqn13}
\end{eqnarray}
and
\begin{eqnarray}
 \delta\ddot{Z} + 3H\delta\dot{Z} - \dot{Z}(k + \dot{\alpha}) - [2\ddot{Z} + 3H\dot{Z}]\alpha = 0
 \label{ref_perturbed_eqn14}
\end{eqnarray}
respectively. However the relation $H_{\mu\nu\rho} =
\epsilon_{\mu\nu\rho\sigma}\partial^{\sigma}Z$, immediately leads
to the following expressions,
\begin{eqnarray}
 h_4\delta h^4 = \dot{Z}\delta\dot{Z} - \frac{3}{a}\dot{Z}^2\delta a\nonumber\\
 h^4\delta h_4 = \dot{Z}\delta\dot{Z} + \frac{3}{a}\dot{Z}^2\delta a\nonumber\\
 \dot{h}_4\delta\dot{h}^4 + \dot{h}^4\delta\dot{h}_4 = 2[\dot{Z}\delta\ddot{Z} + \ddot{Z}\delta\dot{Z}]
 \label{ref_transformation}
\end{eqnarray}
The above relations transform Eqs. (\ref{ref_perturbed_eqn1} -
\ref{ref_perturbed_eqn7}) to Eqs. (\ref{ref_perturbed_eqn8} -
\ref{ref_perturbed_eqn14}). This validates the equivalence between
the set of Eqs. (\ref{ref_perturbed_eqn1} -
\ref{ref_perturbed_eqn7}) and Eqs. (\ref{ref_perturbed_eqn8} -
\ref{ref_perturbed_eqn14}).

Thereby, the first order perturbed cosmological equations are
equivalent for the cases when $H_{\mu\nu\alpha}$ is represented by
a massless scalar field $Z$. In the present context, as we are
interested to determine the spectral index and tensor-to-scalar
ratio, we show the equivalence (between KR and massless scalar
field) only up to first order perturbed equations and we do not
consider the higher order perturbations. However the investigation
of such equivalence for second (or higher) order perturbations is
important, which we hope to address in a future work.

Having discussed the above issues, we shall calculate the
slow-roll indices and the corresponding observational indices for
the KR $F(R)$ gravity model at hand. This is the subject of the
next section.

\section{Inflationary Phenomenology of the Kalb-Ramond $F(R)$ model}

Recall that the original higher curvature $F(R)$ model is given by
the action given in Eq. (\ref{action1}). The spacetime metric for
this $F(R)$ model can be extracted from the corresponding
scalar-tensor theory (see Eq. (\ref{4d metric})) with the help of
inverse conformal transformation. Thus the line element in $F(R)$
model turns out to be,
\begin{eqnarray}
 ds^2&=&e^{\sqrt{\frac{2}{3}}\kappa\xi(t)} \bigg[-dt^2 + a^2(t)\big(dx^2 + dy^2 + dz^2\big)\bigg]\nonumber\\
 &=&-d\tau^2 + s^2(\tau)\big(dx^2 + dy^2 + dz^2\big)
 \label{4d metric II}
\end{eqnarray}
where $\tau(t)$, $s(\tau)$ are the cosmic time and the scale
factor respectively in the $F(R)$ frame and they are defined as
follows:
\begin{eqnarray}
\tau(t) = \int dt
e^{[\frac{1}{2}\sqrt{\frac{2}{3}}\kappa\xi(t)]}\, , \label{cosmic
time1 F(R)}
\end{eqnarray}
and
\begin{eqnarray}
 s(\tau(t)) = e^{[\frac{1}{2}\sqrt{\frac{2}{3}}\kappa\xi(t)]}
 a(t)\, .
 \label{scale factor1 F(R)}
\end{eqnarray}
Clearly Eq. (\ref{cosmic time1 F(R)}) indicates that $\tau(t)$ is
a monotonically increasing function of the cosmic time $t$.
Consequently the Hubble parameter ($H_F$) in the $F(R)$ frame is
defined as $H_F = \frac{1}{s(\tau)}\frac{ds}{d\tau}$. Using Eqs.
(\ref{cosmic time1 F(R)}) and (\ref{scale factor1 F(R)}), we can
relate the Hubble parameter to that of the corresponding
scalar-tensor model as follows,
\begin{eqnarray}
 H_{F} = e^{-\frac{\sqrt{2}\kappa}{2\sqrt{3}}\xi(t)}\bigg[H +
 \frac{\sqrt{2}\kappa}{2\sqrt{3}}\dot{\xi}\bigg]\, ,
  \label{hubble F(R)}
\end{eqnarray}
where $H$ is the Hubble parameter in the scalar-tensor frame and
the ``dot'' denotes differentiation with respect to the cosmic
time.

\subsection{The Slow-roll Case}

\begin{figure}[h]
\centering
\includegraphics[width=18pc]{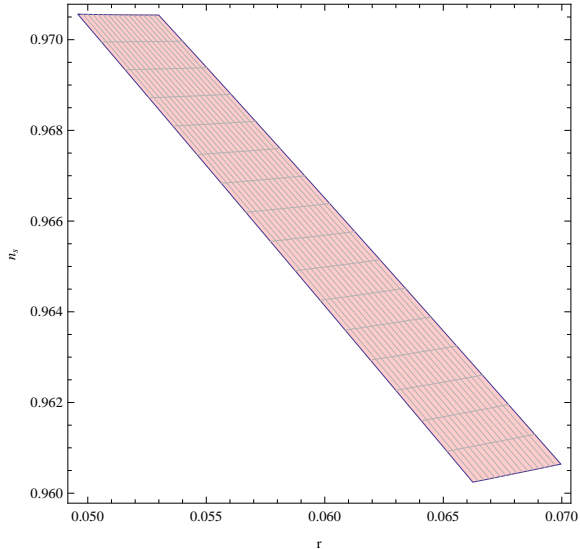}
\caption{{\it{The parametric plot of $n_s$ and $r$ as functions of
$\kappa^2h_0\alpha$ and $\beta/\alpha$, for $ 0\leq
\kappa^2h_0\alpha \leq 10^{-5}$ and $0.22\leq \beta/\alpha\leq
0.26$.}}} \label{plot1}
\end{figure}

\begin{figure}[h]
\centering
\includegraphics[width=18pc]{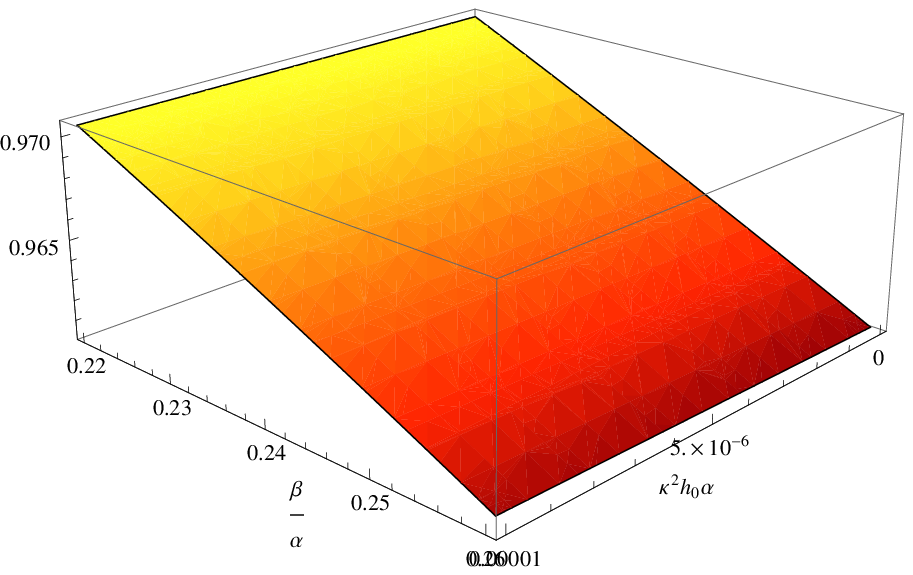}
\includegraphics[width=2pc]{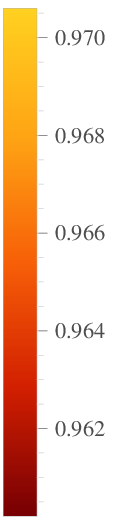}
\caption{{\it{The three dimensional plot of the spectral index
$n_s$ as a function of $\kappa^2h_0\alpha$ and $\beta/\alpha$, for
$ 0\leq \kappa^2h_0\alpha \leq 10^{-5}$ and $0.22\leq
\beta/\alpha\leq 0.26$.}}} \label{plot3}
\end{figure}

Now we shall investigate the phenomenological aspects of the
modified gravity model at hand, and we shall calculate in detail
the spectral index of the primordial curvature perturbations and
the tensor-to-scalar ratio. The first slow-roll parameter reads,
 \begin{eqnarray}
  \epsilon_{F} = -\frac{1}{H_F^2}\frac{dH_F}{d\tau}
  \label{slow roll parameter F(R)1}
 \end{eqnarray}
 where $H_F$ is the Hubble parameter in the $F(R)$ model and it is defined in Eq. (\ref{hubble F(R)}). In order to find the explicit expression of $\epsilon_F$,
 we calculate $\frac{dH_F}{d\tau}$ by differentiating Eq. (\ref{hubble F(R)}) with respect to the conformal time $\tau$ in this frame,
 \begin{eqnarray}
  \frac{dH_{F}}{d\tau} = e^{-\frac{\sqrt{2}\kappa}{\sqrt{3}}\xi(t)}\bigg[\dot{H} -
  \frac{\sqrt{2}\kappa}{2\sqrt{3}}H\dot{\xi}\bigg]\, ,
  \label{hubble derivative F(R)}
 \end{eqnarray}
where we used the slow-roll approximation condition. With the
expressions of $H_F$ and $\frac{dH_F}{d\tau}$ at hand, along with
the slow-roll field equations, we obtain the following form of
$\epsilon_F$,
 \begin{eqnarray}
  \epsilon_{F} = \bigg[\frac{\frac{3\kappa^2h_0}{2} - \frac{\kappa}{\sqrt{6}}V'(\xi)}
  {\kappa^2V(\xi) + \frac{\kappa^2h_0}{2} - \frac{\kappa}{\sqrt{6}}V'(\xi)}\bigg]
  \label{slow roll parameter F(R)2}
 \end{eqnarray}
where $V(\xi)$ and its derivative are given in Eqs.
(\ref{potential}) and (\ref{first derivative}) respectively.  As
we mentioned earlier, the second rank antisymmetric KR field can
be equivalently expressed as a vector field which can be further
recast as a derivative of a massless scalar field (see the
Appendix). As a consequence, the spectral index and
tensor-to-scalar ratio in the present context are defined as
follows
\cite{Noh:2001ia,Hwang:2005hb,Hwang:2002fp,Kaiser:2013sna}:
 \begin{eqnarray}
  n_s = \big[1 - 4\epsilon_F - 2\epsilon_2 + 2\epsilon_3 -
  2\epsilon_4\big]\bigg|_{\tau_0}\, ,
  \label{spectral index1}
 \end{eqnarray}
 and
 \begin{eqnarray}
  r = 8\kappa^2 \frac{\varTheta}{F'(R)}\bigg|_{\tau_0}\, ,
  \label{ratio1}
 \end{eqnarray}
respectively. The slow-roll parameters ($\epsilon_F$,
$\epsilon_2$, $\epsilon_3$, $\epsilon_4$) are defined as follows,
\begin{eqnarray}
 \epsilon_F&=&-\frac{1}{H_F^2} \frac{dH_F}{d\tau},\,\,\,\,\,\,\epsilon_2 = \frac{1}{2\rho_{KR}H_F} \frac{d\rho_{KR}}{d\tau}\, ,\nonumber\\
 \epsilon_3&=&\frac{1}{2F'(R)H_F} \frac{d F'(R)}{d\tau},\,\,\,\,\,\,\epsilon_4 = \frac{1}{2 E H_F}
 \frac{d E}{d\tau}\, ,
 \label{various slow roll parameters}
\end{eqnarray}
where
\begin{eqnarray}
 E = \frac{\varTheta F'(R)H_F^2}{\rho_{KR}}\, ,
 \label{E}
\end{eqnarray}
and $\varTheta$ is,
\begin{eqnarray}
 \varTheta = \frac{\rho_{KR}}{F'(R)H_F^2}\bigg[F'(R) +
 \frac{3}{2\kappa^2\rho_{KR}}\bigg(\frac{d}{d\tau}F'(R)\bigg)^2\bigg]\,
 .
 \label{vartheta}
\end{eqnarray}
In the equations above, $\rho_{KR}$  is ($= H_{123}H^{123}$) the
energy density of the KR field in the $F(R)$ gravity model.
However due to Eqs. (\ref{redefine}) and (\ref{solution of KR
energy density}), the variation of $\rho_{KR}$ yields $\rho_{KR} =
e^{-2\sqrt{\frac{2}{3}}\kappa\xi(t)} \frac{h_0}{2a^6}$. Keeping
this in mind, now we are going to determine the explicit
expressions of various terms appearing in the right hand side of
Eqs. (\ref{spectral index1}) and (\ref{ratio1}). Let us start with
$\epsilon_F$, which is given by,
 \begin{eqnarray}
  \epsilon_{F} = \bigg[\frac{\frac{3\kappa^2h_0}{2} - \frac{\kappa}{\sqrt{6}}V'(\xi)}
  {\kappa^2V(\xi) + \frac{\kappa^2h_0}{2} -
  \frac{\kappa}{\sqrt{6}}V'(\xi)}\bigg]\, .
  \label{first slow roll parameter}
 \end{eqnarray}
With regard to $\epsilon_2$, as it is mentioned in Eq.
(\ref{various slow roll parameters}), this parameter $\epsilon_2$
 is related with the variation of KR field energy density and thus
 its calculation requires the field equation of the KR field. However the KR field energy
density in $F(R)$, namely $\rho_{KR}$, and in the corresponding
scalar-tensor theory, namely $\tilde{\rho}_{KR}$, are connected by
$\rho_{KR} = e^{-2\sigma}\tilde{\rho}_{KR}$ (with $\sigma =
\sqrt{\frac{2}{3}}\kappa\xi$). Differentiating both sides of this
expression, with respect to the $F(R)$ frame conformal cosmic time
$\tau$, one gets,
\begin{eqnarray}
  \frac{d\rho_{KR}}{d\tau}&=&\frac{d}{dt}\big[e^{-2\sigma}\tilde{\rho}_{KR}\big] \frac{dt}{d\tau}\nonumber\\
  &=&\rho_{KR} e^{-\sigma/2}\bigg[\frac{1}{\tilde{\rho}_{KR}}\frac{d\tilde{\rho}_{KR}}{dt} -
  2\frac{d\sigma}{dt}\bigg]\, ,
  \label{time variation 1}
 \end{eqnarray}
where we used the relation between $\tau$ and $t$ given in Eq.
(\ref{cosmic time1 F(R)}). Recall that the evolution of
$\tilde{\rho}_{KR}$ (see Eq. (\ref{evolution of KR energy
density})) is,
 \begin{eqnarray}
  \frac{1}{\tilde{\rho}_{KR}}\frac{d\tilde{\rho}_{KR}}{dt} + 6H =
  0\, .
  \label{time variation 2}
 \end{eqnarray}
 With the help of expression (\ref{time variation 1}), the above equation can be written in terms of $\rho_{KR}$ as follows,
 \begin{eqnarray}
  \frac{\rho_{KR}'}{2\rho_{KR}} + e^{-\sigma/2} \dot{\sigma} + 3e^{-\sigma/2}H =
  0\, ,
  \label{time variation 3}
 \end{eqnarray}
where ``prime'' and ``dot'' represent the derivatives with respect
to $\tau$ and $t$ respectively. Combining Eqs. (\ref{time
variation 3})  and (\ref{hubble F(R)}), we obtain the final form
of $\epsilon_2$ which is,
\begin{eqnarray}
 \epsilon_2&=&\frac{\rho_{KR}'}{2\rho_{KR}H_F}\nonumber\\
 &=&-3 + \frac{\dot{\sigma}}{2H_F}e^{-\sigma/2}\, .
 \label{second slow roll parameter}
\end{eqnarray}
Now let us turn our focus on the calculation of $\epsilon_3$, so
by using $F(R) = R + \alpha R^2 + \beta R^2\ln{(\beta R)}$,
$\epsilon_3$ can be simplified as,
\begin{eqnarray}
 \epsilon_3&=&\frac{1}{2F'(R)H_F}\frac{dF'(R)}{d\tau}\nonumber\\
 &=&\frac{1}{2H_FR\big(2\alpha + \beta + 2\beta\ln{(\beta R)}\big)}\frac{d}{d\tau}\bigg[\big(2\alpha + \beta + 2\beta\ln{(\beta
 R)}\big)R\bigg]\, ,
 \label{third 1}
\end{eqnarray}
where we consider $1 + (2\alpha + \beta)R \simeq (2\alpha +
\beta)R$ near the time when the cosmological perturbations exit
the horizon (as $n_s$ and $r$ are eventually calculated at the
time of horizon crossing). Furthermore, at the time of the horizon
crossing, the spacetime curvature is large (compared to the
present one). In such a large curvature regime, the variation of
$\ln{(R)}$ becomes small and thus one may safely write $\ln{(R)}
\simeq \ln{(R_0)}$, where $R_0$ is the spacetime curvature at the
time of horizon crossing. This consideration leads to the
following form of $\frac{dF'(R)}{d\tau}$,
\begin{eqnarray}
 \frac{dF'(R)}{d\tau} = \big[2\alpha + \beta + 2\beta\ln{(\beta R_0)}\big]\frac{dR}{d\tau}
 \nonumber
\end{eqnarray}
Moreover for the FRW metric, the Ricci scalar takes the form
$R\simeq 12H_F^2$. Using this expression, we get the final form of
$\epsilon_3$ which is,
\begin{eqnarray}
 \epsilon_3&=&\frac{1}{H_F^2}\frac{dH_F}{d\tau} = -\epsilon_F\nonumber\\
 &=&-\bigg[\frac{\frac{3\kappa^2h_0}{2} - \frac{\kappa}{\sqrt{6}}V'(\xi)}
  {\kappa^2V(\xi) + \frac{\kappa^2h_0}{2} - \frac{\kappa}{\sqrt{6}}V'(\xi)}\bigg]
 \label{third slow roll paramater}
\end{eqnarray}
With regard to  $\epsilon_4$, we need to calculate the function
$E$ which is defined as $E = \frac{\varTheta
F'(R)H_F^2}{\rho_{KR}}$. By differentiating this expression with
respect to $\tau$, we obtain,
\begin{eqnarray}
 \frac{E'}{EH_F} = \frac{\varTheta'}{\varTheta H_F} + \frac{1}{F'(R)H_F}\frac{dF'(R)}{d\tau} + 2\frac{H_F'}{H_F^2} -
 \frac{\rho_{KR}'}{\rho_{KR}H_F}\, .
 \label{fourth 1}
\end{eqnarray}
The above expression can be further simplified by using Eqs.
 (\ref{second slow roll parameter}) and (\ref{third slow roll paramater}), so we get,
\begin{eqnarray}
 \frac{E'}{EH_F} = \frac{\varTheta'}{\varTheta H_F} - 4\epsilon_F + 6 -
 \frac{\dot{\sigma}}{H_F}e^{-\sigma/2}\, .
 \label{fourth 2}
\end{eqnarray}
At this point, what remains is to determine $\varTheta$ in order
to get the final expression of $\epsilon_4$ as well as of $n_s$.
The function $\varTheta$ is given by,
\begin{eqnarray}
 \varTheta&=&\frac{\rho_{KR}}{H_F^2} + \frac{3}{2\kappa^2F'(R)H_F^2}\bigg(\frac{dF'(R)}{d\tau}\bigg)^2\nonumber\\
 &=&\frac{\rho_{KR}}{H_F^2} +
 6\epsilon_F^2\big(\frac{F'(R)}{\kappa^2}\big)\, .
 \label{fourth 3}
\end{eqnarray}
Differentiation of both sides of Eq. (\ref{fourth 3}) yields the
following expression,
\begin{eqnarray}
 \frac{\varTheta'}{\varTheta H_F} = -2\epsilon_F + 2\frac{\epsilon_F'}{\epsilon_F H_F} + \frac{\kappa^2\rho_{KR}}{6F'(R)\epsilon_F^2 H_F^3}
 \bigg[-6H_F + e^{-\sigma/2}\dot{\sigma} -
 \frac{H_F'}{H_F}\bigg]\, ,
 \label{fourth 4}
\end{eqnarray}
where we used Eqs. (\ref{second slow roll parameter}) and
(\ref{third slow roll paramater}). However the above expression
together with Eq. (\ref{fourth 2}) leads to the final form of
$\epsilon_4$, which is,
\begin{eqnarray}
 \epsilon_4&=&-3\epsilon_F + \frac{\epsilon_F'}{\epsilon_F H_F} + 3 - \frac{\dot{\sigma}}{2H_F}e^{-\sigma/2}\nonumber\\
 &+&\frac{\kappa^2\rho_{KR}}{6F'(R)\epsilon_F^2 H_F^3} \bigg[-6H_F + e^{-\sigma/2}\dot{\sigma} -
 \frac{H_F'}{H_F}\bigg]\, .
 \label{fourth slow roll parameter}
\end{eqnarray}
Having found the slow-roll indices, we can now calculate the
spectral index $n_s$, by using the above expressions for the
slow-roll indices, so finally we get,
\begin{eqnarray}
 n_s = 1 - 2\frac{\epsilon_F'}{\epsilon_F H_F}
 + \frac{\kappa^2\rho_{KR}}{6F'(R)\epsilon_F^2 H_F^3} \bigg[-6H_F + e^{-\sigma/2}\dot{\sigma} -
 \frac{H_F'}{H_F}\bigg]\, .
 \label{spectral index intermediate}
\end{eqnarray}
As it can be seen, in the absence of KR field, which can be
obtained by setting $\rho_{KR}=0$, $n_s$ takes the form $n_s = 1 -
2\frac{\epsilon_F'}{\epsilon_F H_F}$, which is in agreement with
the expression of spectral index in a pure vacuum $F(R)$ gravity
model \cite{Noh:2001ia,Hwang:2005hb,Hwang:2002fp,Kaiser:2013sna}.
However due to the presence of the KR field, $n_s$ gets modified
by the terms proportional to $\rho_{KR}$. Taking these
modifications into account, the final form of $n_s$ given in Eq.
(\ref{spectral index intermediate}) can be cast as follows,
\begin{eqnarray}
 n_s = 1&-&2\bigg[\frac{-\frac{\kappa^2h_0}{2} + \frac{1}{72}\bigg(\frac{\sqrt{2/3}\kappa V'(\xi)V''(\xi)}
 {\kappa^2 V(\xi) + \frac{\kappa^2h_0}{2}}\bigg)}{\frac{3\kappa^2h_0}{2} - \sqrt{1/6} \kappa V'(\xi)}\bigg]
 + 2\bigg[\frac{-3\kappa^2h_0 - \frac{1}{72}\bigg(\frac{\big(\sqrt{2/3}\kappa V'(\xi)\big)^2}
 {\kappa^2 V(\xi) + \frac{\kappa^2h_0}{2}}\bigg)}{\frac{\kappa^2h_0}{2} + \kappa^2 V(\xi)
 - \sqrt{2/3}\kappa V'(\xi)}\bigg]\nonumber\\
 &+&\frac{\kappa^2h_0}{12\big[2\alpha + \beta + 2\beta\ln{(\beta R_0)}\big]\big[\frac{\kappa^2h_0}{2} - \frac{1}{6}\sqrt{2/3} \kappa V'(\xi)\big]^2}
 \nonumber\\
 &-&\frac{\kappa^2h_0}{12\big[2\alpha + \beta + 2\beta\ln{(\beta R_0)}\big]\big[\frac{\kappa^2h_0}{2} - \frac{1}{6}\sqrt{2/3} \kappa V'(\xi)\big]
 \big[\kappa^2V(\xi) + \frac{\kappa^2h_0}{2} - \sqrt{2/3} \kappa
 V'(\xi)\big]}\nonumber\, ,
 \label{spectral index2}
\end{eqnarray}
where we used the slow-roll field equations  (\ref{slow roll
equation1a}) and (\ref{slow roll equation2a}). Accordingly, we can
calculate the tensor-to-scalar ratio $r$, which is defined as $r =
8\kappa^2 \frac{\varTheta}{F'(R)}$. By using the explicit
expression of $\varTheta$ given in  Eq. (\ref{fourth 3}), we
obtain,
\begin{eqnarray}
 r = 8\kappa^2 \frac{\rho_{KR}}{H_F^2 F'(R)} + 48 \bigg(\frac{1}{2H_F
 F'(R)}\frac{dF'(R)}{d\tau}\bigg)^2\bigg|_{\tau_0}\, .
 \label{fifth 1}
\end{eqnarray}
With $F'(R) = 1 + [2\alpha + \beta + 2\beta\ln{(\beta R)}]R \simeq
[2\alpha + \beta + 2\beta\ln{(\beta R)}]R$
 along with the expressions $R \simeq 12H_F^2$ and $H_F = e^{-\sigma/2}\big[H + \frac{\dot{\sigma}}{2}$
 (see Eq. (\ref{hubble F(R)})), one gets,
 \begin{eqnarray}
  8\kappa^2 \frac{\rho_{KR}}{H_F^2 F'(R)}\bigg|_{\tau_0} = \frac{2\kappa^2h_0}{3\big[2\alpha + \beta + 2\beta\ln{(\beta R_0)}\big]
  \bigg[H^2 + H\dot{\sigma}\bigg]^2}\, .
  \label{fifth 2}
 \end{eqnarray}
By using Eqs. (\ref{slow roll equation1a}) and (\ref{slow roll
equation2a}), the above expression can be further simplified to,
 \begin{eqnarray}
  8\kappa^2 \frac{\rho_{KR}}{H_F^2 F'(R)}\bigg|_{\tau_0} = \frac{\frac{2\kappa^2h_0}{3\big[2\alpha + \beta + 2\beta\ln{(\beta R_0)}\big]}}
  {\big[\frac{\kappa^2 V(\xi)}{3} + \frac{\kappa^2h_0}{6} - \frac{1}{3}\sqrt{2/3} \kappa V'(\xi)\big]^2}
  \label{fifth 3}
 \end{eqnarray}
Accordingly, the second term in the right hand side of Eq.
(\ref{fifth 1}) is equal to $48\epsilon_F^2$. Combining the above
expressions, we may obtain the final expression of the
tensor-to-scalar ratio, which is equal to,
\begin{eqnarray}
 r = \frac{\frac{2\kappa^2h_0}{3\big[2\alpha + \beta + 2\beta\ln{(\beta R_0)}\big]}}
  {\big[\frac{\kappa^2 V(\xi)}{3} + \frac{\kappa^2h_0}{6} - \frac{1}{3}\sqrt{2/3} \kappa V'(\xi)\big]^2} +
  48\epsilon_F^2\, .
  \label{ratio intermediate}
\end{eqnarray}
It may be noticed from Eq. (\ref{ratio intermediate}) that for
$\rho_{KR} = 0$ (or equivalently $h_0 = 0$ i.e without the KR
field), $r$ goes to $48\epsilon_F^2$ which is the expression for
the tensor-to-scalar ratio in a pure vacuum $F(R)$ gravity model
\cite{Noh:2001ia,Hwang:2005hb,Hwang:2002fp,Kaiser:2013sna}.
However taking the effect of the KR field into account, and
plugging back the expression of $\epsilon_F$ (obtained in Eq.
(\ref{slow roll parameter F(R)2})) in Eq. (\ref{ratio
intermediate}), we get the final form of $r$ which is,
 \begin{eqnarray}
  r = \frac{\frac{2\kappa^2h_0}{3\big[2\alpha + \beta + 2\beta\ln{(\beta R_0)}\big]}
  + 48\big[\frac{\kappa^2h_0}{2} - \frac{1}{6}\sqrt{2/3}\kappa V'(\xi)\big]^2}
  {\big[\frac{\kappa^2 V(\xi)}{3} + \frac{\kappa^2h_0}{6} - \frac{1}{3}\sqrt{2/3} \kappa
  V'(\xi)\big]^2}\, .
  \label{ratio2}
 \end{eqnarray}
 Thereby the final expressions of $n_s$ and $r$ are given in Eqs. (\ref{spectral index2}) and (\ref{ratio2}) respectively, from which, it is evident
 that both the observational quantities depend on the parameters $\alpha$, $\beta$, $h_0$ and $R_0$. However
 if we consider the values restricted as $\alpha \neq \beta \sim 1/R_0$, then both $n_s$ and $r$ depend on two dimensionless parameters
 $\kappa^h_0\alpha$ and $\beta/\alpha$. The latest observational
 data coming from the Planck 2018 and BICEP2/Keck-Array data
 constrain $n_s$ and $r$ as follows, $n_s = 0.9650 \pm 0.0066$ and $r <
 0.07$. The compatibility with the observational data can be
 achieved if  $0.20 < \frac{\beta}{\alpha} < 0.25$ and $\kappa^2h_0\alpha\bigg|_{max} =
 10^{-5}$. The results are summarized in Table \ref{Table-1}.
\begin{table}[h]
 \centering
  \begin{tabular}{|c| c|}
   \hline \hline
   Parameters & Estimated values\\
   \hline
   $n_s$ & $\simeq 0.9630$\\
   $r$ & $\simeq 0.03$\\
   \hline
  \end{tabular}%
  \caption{Estimated values of various quantities for $\kappa^2h_0\alpha = 5\times 10^{-6}$, $\beta/\alpha = 0.24$}
  \label{Table-1}
 \end{table}
Finally, by taking $\alpha = 1$ (in Planck units) we obtain
$h_0^{max} \sim 10^{71}$ (GeV)$^{4}$. Therefore the present model
along with the constraints of $Planck$ 2018 gives an upper bound
on the KR field energy density during the primordial era of our
Universe, which is $\big(\rho_{KR}\big)^{max} \sim 10^{71}$
(GeV)$^{4}$. We can also see that the spectral index and the
tensor-to-scalar ratio are simultaneously compatible with the
observational constraints by looking Fig. \ref{plot1} where we
present the parametric plot of $n_s$ and $r$ as functions of
$\kappa^2h_0\alpha$ and $\beta/\alpha$, for $ 0\leq
\kappa^2h_0\alpha \leq 10^{-5}$ and $0.22\leq \beta/\alpha\leq
0.26$. As it can be seen from Fig. \ref{plot1}, there exist a wide
range of the free parameters for which the simultaneous
compatibility of the spectral index and of the tensor-to-scalar
ratio can be achieved.
This can also be seen in Fig. \ref{plot3} where we present the
three dimensional plot of the spectral index $n_s$ as a function
of $\kappa^2h_0\alpha$ and $\beta/\alpha$, for $ 0\leq
\kappa^2h_0\alpha \leq 10^{-5}$ and $0.22\leq \beta/\alpha\leq
0.26$.

In order to better understand the effect of logarithmic correction
over $R^2$ Starobinsky inflation model, and the actual impact of
the KR field on the inflationary phenomenology, let us here
discuss the observational predictions of the logarithmic $R^2$
model in the presence or not of the KR field, and directly compare
the results to the standard $R^2$ models for the same values of
the relevant free parameters.

Let us start with the standard $R^2$ Starobinsky inflation model
in the absence of the KR field, with the functional form of the
$F(R)$ gravity being in this case $R + R^2/m^2$. The constraints
for the free parameters are  $0.02 \lesssim m^2/R_0 \lesssim
0.07$, where $R_0$ is the value of the the Ricci scalar at the
time of horizon crossing. For these values we obtain $0.960 \leq
n_s \leq 0.970$ and $r < 0.007$. With regard to the $R^2$ model in
the presence of a Kalb-Ramond field, the constraints are $0.02
\lesssim m^2/R_0 \lesssim 0.07$ and $\frac{\kappa^2h_0}{m^2}
\lesssim 3\times10^{-3}$, and it can be shown that $n_s$ takes
values in the range $0.95\leq n_s \leq 1$, while $r < 0.060$. As
for the logarithmic corrected $F(R)=R + R^2/m^2 + \beta
R^2\ln{(\beta R)}$ gravity without KR field, for $0.40 \lesssim
m^2/R_0 \lesssim 0.50$ for $\beta = \frac{1}{2R_0}$, it can be
shown that $n_s$ takes values in the range $0.96\leq n_s \leq
0.97$, while $r < 0.060$. Finally, for the logarithmic corrected
$R^2$ model with KR field, for $0.40 \lesssim m^2/R_0 \lesssim
0.50$, $\frac{\kappa^2h_0}{m^2} \lesssim 10^{-5}$ and $\beta =
\frac{1}{2R_0}$, it can be shown that $n_s$ takes values in the
range $0.96\leq n_s \leq 0.97$, while $r < 0.065$. Thus the effect
of the KR field is to increase the amount of gravitational
radiation predicted from the standard $R^2$ Starobinsky inflation
model.

\subsection{The Constant-roll Case}

In the case that the constant-roll approximation is considered,
slow-roll indices are given by the following expressions,
\cite{Noh:2001ia,Hwang:2005hb,Hwang:2002fp,Kaiser:2013sna},
\begin{eqnarray}
 \epsilon_F&=&-\frac{1}{H_F^2} \frac{dH_F}{d\tau},\,\,\,\,\,\epsilon_2 = \frac{1}{2\rho_{KR}H_F} \frac{d\rho_{KR}}{d\tau}\nonumber\\
 \epsilon_3&=&\frac{1}{2F'(R)H_F} \frac{dF'(R)}{d\tau},\,\,\,\,\,\epsilon_4 = \frac{1}{2EH_F}
 \frac{dE}{d\tau}\, ,
 \label{various slow roll parameters}
\end{eqnarray}
and in this case, the function $E$ is equal to,
\begin{eqnarray}
 E = \frac{\varTheta F'(R)H_F^2}{\rho_{KR}}\, ,
 \label{E}
\end{eqnarray}
where $\varTheta$ in this case, is equal to,
\begin{eqnarray}
 \varTheta = \frac{\rho_{KR}}{(1 + \epsilon_3)^2F'(R)H_F^2}\bigg[F'(R) +
 \frac{3}{2\kappa^2\rho_{KR}}\bigg(\frac{d}{d\tau}F'(R)\bigg)^2\bigg]\,
 .
 \label{vartheta}
\end{eqnarray}
Recall that $\rho_{KR}$ ($= H_{123}H^{123}$) denotes the energy
density of the KR field in the $F(R)$ gravity model. It can easily
be seen that the definitions of the slow-roll indices $\epsilon_i$
are the same in comparison to those corresponding to the slow-roll
condition, except for the factor $(1+\epsilon_3)^2$ in the
denominator of $\varTheta$, however, as we will see, the resulting
functional forms in terms of the model parameters $h_0$, $\alpha$,
$\beta$, will actually change due to the constant-roll condition.
The observational indices in terms of the slow-roll indices are
equal to
\cite{Noh:2001ia,Hwang:2005hb,Hwang:2002fp,Kaiser:2013sna},
\begin{eqnarray}
 n_s = 4 - 2\sqrt{\bigg[\frac{1}{4} + \frac{\big(1+\epsilon_F+\epsilon_2-\epsilon_3+\epsilon_4\big)\big(2+\epsilon_2-\epsilon_3+\epsilon_4\big)}
 {(1 - \epsilon_F)^2}\bigg]}
 \label{spectral_constant1}
\end{eqnarray}
and
\begin{eqnarray}
 r&=&8\kappa^2 \frac{\varTheta}{F'(R)}\nonumber\\
 &=&\frac{1}{(1+\epsilon_3)^2} \bigg[8\kappa^2\frac{\rho_{KR}}{F'(R)H_F^2} + \frac{12}{F'(R)^2H_F^2}\big(\frac{dF'(R)}{d\tau}\big)^2\bigg]
 \label{ratio_constant1}
\end{eqnarray}
Let us find the explicit form of each slow-roll index, so we start
off with the calculation of $\epsilon_F$, and by using the Hubble
rate of Eq. (\ref{hubble F(R)}), the Hubble parameter in the
$F(R)$ frame ($H_F$) and in the scalar-tensor frame ($H$) are
related by $H_{F} =
e^{-\frac{\sqrt{2}\kappa}{2\sqrt{3}}\xi(t)}\bigg[H +
\frac{\sqrt{2}\kappa}{2\sqrt{3}}\dot{\xi}\bigg]$. Differentiating
with respect to $\tau$ leads to the following expression,
 \begin{eqnarray}
  \frac{dH_F}{d\tau}&=&e^{-\sqrt{\frac{2}{3}}\kappa\xi(t)} \bigg[\dot{H} + \frac{1}{2}\sqrt{\frac{2}{3}}\kappa\ddot{\xi}
  - \frac{1}{2}\sqrt{\frac{2}{3}}\kappa H\dot{\xi} - \frac{1}{6}\kappa^2\dot{\xi}^2\bigg]\nonumber\\
  &=&e^{-\sqrt{\frac{2}{3}}\kappa\xi(t)} \bigg[\dot{H} - \frac{1}{2}\sqrt{\frac{2}{3}}\kappa H\dot{\xi}(1-\gamma)
  - \frac{\kappa^2}{6}A^2a^{2\gamma}\bigg]\, ,
  \label{hubble derivative constant}
 \end{eqnarray}
where the ``dot'' denotes differentiation with respect to $t$, and
in the second line we used the constant-roll condition. Using
these expressions for $H_F$ and $H_F'(\tau)$, we obtain the
explicit form of $\epsilon_F$ in terms of the model parameters as
follows,
 \begin{eqnarray}
  \epsilon_F\bigg|_{\tau_0} = \bigg[\frac{\frac{3\kappa^2h_0}{2} - \frac{\kappa(1-\gamma)}{2(1+\gamma/3)}\sqrt{\frac{2}{3}}V'(\xi) + 2\kappa^2A^2a_0^{2\gamma}}
  {\kappa^2A^2a_0^{2\gamma} + \kappa^2V(\xi) + \frac{\kappa^2h_0}{2} - \frac{\kappa}{(1+\gamma/3)}\sqrt{\frac{2}{3}}V'(\xi)}\bigg]
  \label{first parameter constant}
 \end{eqnarray}
where $a_0 = a(\tau_0)$ and $V(\xi)$ and its derivative are given
in Eqs. (\ref{potential}) and (\ref{first derivative})
respectively. By comparing Eqs. (\ref{first slow roll parameter})
and (\ref{first parameter constant}) we may conclude that
$\epsilon_F = -H_F'(\tau)/H_F^2$  differs in the slow-roll and
constant-roll cases, because of the presence of $\gamma$ and $A$,
as expected. Let us now turn our focus on the slow-roll parameters
$\epsilon_2$ and $\epsilon_3$, and by using similar
considerations, we obtain the following expressions, namely,
 \begin{eqnarray}
  \epsilon_2 = -3 + e^{-[\frac{1}{2}\sqrt{\frac{2}{3}}\kappa\xi]}\frac{1}{2H_F}\sqrt{\frac{2}{3}}\kappa\dot{\xi}
  \label{second parameter constant}
 \end{eqnarray}
 and
 \begin{eqnarray}
  \epsilon_3&=&-\epsilon_F\nonumber\\
  &=&-\bigg[\frac{\frac{3\kappa^2h_0}{2} - \frac{\kappa(1-\gamma)}{2(1+\gamma/3)}\sqrt{\frac{2}{3}}V'(\xi) + 2\kappa^2A^2a_0^{2\gamma}}
  {\kappa^2A^2a_0^{2\gamma} + \kappa^2V(\xi) + \frac{\kappa^2h_0}{2} -
  \frac{\kappa}{(1+\gamma/3)}\sqrt{\frac{2}{3}}V'(\xi)}\bigg]\, .
  \label{third parameter constant}
 \end{eqnarray}
Finally, in order to obtain $\epsilon_4$, recall that $E$ is
defined as $E = \frac{\varTheta
F'(R)H_F^2}{\rho_{KR}}(1+\epsilon_3)^2$. Differentiating with
respect to $\tau$, we get,
\begin{eqnarray}
 \epsilon_4&=&\frac{E'}{2EH_F}\nonumber\\
 &=&\frac{\varTheta'}{2\varTheta H_F} + \frac{1}{2H_F F'(R)}\frac{dF'(R)}{d\tau} + \frac{H_F'}{H_F^2} - \frac{\rho_{KR}'}{2H_F\rho_{KR}}
 + \frac{\epsilon_3'}{H_F(1+\epsilon_3)}
 \label{fourth parameter constant 1}
\end{eqnarray}
Using Eqs. (\ref{second parameter constant})and (\ref{third
parameter constant}), the above expression can be simplified
$\epsilon_4$ takes the following form,
\begin{eqnarray}
 \epsilon_4 = \frac{\varTheta'}{2\varTheta H_F} - 2\epsilon_F - \epsilon_2 +
 \frac{\epsilon_3'}{H_F(1+\epsilon_3)}\, .
 \label{fourth parameter constant 2}
\end{eqnarray}
The remaining part is to determine $\varTheta$ which is defined
as,
\begin{eqnarray}
 \varTheta&=&\frac{1}{(1+\epsilon_3)^2} \bigg[\frac{\rho_{KR}}{H_F^2} + \frac{3}{2\kappa^2F'(R)H_F^2}\bigg(\frac{dF'(R)}{d\tau}\bigg)^2\bigg]\nonumber\\
 &=&\frac{1}{(1+\epsilon_3)^2} \bigg[\frac{\rho_{KR}}{H_F^2} + \frac{6F'(R)\epsilon_F^2}{\kappa^2}\bigg]
 \label{fourth parameter constant 3}
\end{eqnarray}
By differentiating Eq. (\ref{fourth parameter constant 3}) and
after some algebra, we obtain,
\begin{eqnarray}
 \frac{\varTheta'}{2\varTheta H_F} + \frac{\epsilon_3'}{H_F(1+\epsilon_3)} = -\epsilon_F + \frac{\epsilon_F'}{\epsilon_F H_F}
 + \frac{\kappa^2}{12\epsilon_F^2 H_F F'(R)}
 \frac{d}{d\tau}\big[\frac{\rho_{KR}}{H_F^2}\big]\, .
 \label{fourth parameter constant 4}
\end{eqnarray}
The evolution of $\tilde{\rho}_{KR}$ (see Eq. (\ref{solution of KR
energy density})) along with the conformal transformation give the
variation of KR field energy density in the $F(R)$ frame, which
is, $\rho_{KR} = e^{-2\sqrt{\frac{2}{3}}\kappa\xi}
\frac{h_0}{a^6}$. Using this expression, Eq. (\ref{fourth
parameter constant 4}) can be rewritten as follows,
\begin{eqnarray}
 \frac{\varTheta'}{2\varTheta H_F} + \frac{\epsilon_3'}{H_F(1+\epsilon_3)}&=&-\epsilon_F + \frac{\epsilon_F'}{\epsilon_F H_F}\nonumber\\
 &+&\frac{\kappa^2\rho_{KR}}{12\epsilon_F^2 H_F^3 F'(R)}\bigg[-6H_F + e^{-[\frac{1}{2}\sqrt{\frac{2}{3}}\kappa\xi]} \sqrt{\frac{2}{3}}\kappa\dot{\xi}
 - \frac{2H_F'}{H_F}\bigg]\, .
 \label{fourth parameter constant 5}
\end{eqnarray}
Plugging back the above expression into Eq. (\ref{fourth parameter
constant 2}), we obtain the final form of $\epsilon_4$, which is,
\begin{eqnarray}
 \epsilon_4&=&-\epsilon_2 - 3\epsilon_F + \frac{\epsilon_F'}{\epsilon_F H_F}\nonumber\\
 &+&\frac{\kappa^2\rho_{KR}}{12\epsilon_F^2 H_F^3 F'(R)}\bigg[-6H_F + e^{-[\frac{1}{2}\sqrt{\frac{2}{3}}\kappa\xi]} \sqrt{\frac{2}{3}}\kappa\dot{\xi}
 - \frac{2H_F'}{H_F}\bigg]\, .
 \label{fourth parameter constant}
\end{eqnarray}
In view of the above resulting expressions, the spectral index and
the tensor-to-scalar ratio in the constant-roll approximation
read,
\begin{eqnarray}
 n_s = 4 - 2\sqrt{\frac{1}{4} + \frac{D_1D_2}{D_3}}\, ,
 \label{spectral index constant}
\end{eqnarray}
and
\begin{eqnarray}
 r = \frac{D_4}{(1+\epsilon_3)^2}\, ,
 \label{ratio constant}
\end{eqnarray}
where the functions $D_i$ are defined as follows,
\begin{eqnarray}
 &D_1&= 1 + \epsilon_F + \epsilon_2 - \epsilon_3 + \epsilon_4\nonumber\\
&=&1 - \bigg[\frac{\frac{3\kappa^2h_0}{2} -
\frac{\kappa(1-\gamma)}{2(1+\gamma/3)}\sqrt{\frac{2}{3}}V'(\xi) +
2\kappa^2A^2a_0^{2\gamma}}
  {\kappa^2A^2a_0^{2\gamma} + \kappa^2V(\xi) + \frac{\kappa^2h_0}{2} - \frac{\kappa}{(1+\gamma/3)}\sqrt{\frac{2}{3}}V'(\xi)}\bigg]
  + \bigg[\frac{-\frac{\kappa^2h_0}{2} + \frac{1}{72}\bigg(\frac{\sqrt{2/3}\frac{\kappa}{(1+\gamma/3)^2}V'(\xi)V''(\xi)}{\kappa^2 V(\xi)+\kappa^2h_0/2}\bigg)}
  {\frac{3\kappa^2h_0}{2} - \frac{\kappa(1-\gamma)}{2(1+\gamma/3)}\sqrt{2/3} V'(\xi) + 2\kappa^2A^2}\bigg]\nonumber\\
  &+&\bigg[\frac{\frac{3\kappa^2h_0}{2} + \frac{1}{72}\bigg(\frac{\big(\sqrt{2/3}\frac{\kappa}{(1+\gamma/3)}V'(\xi)\big)^2}
  {\kappa^2 V(\xi)+\kappa^2h_0/2}\bigg)}
  {\frac{\kappa^2h_0}{2} - \frac{\kappa}{(1+\gamma/3)}\sqrt{2/3} V'(\xi) + \kappa^2 V(\xi) + \kappa^2 A^2}\bigg]\nonumber\\
  &-&\bigg[\frac{3\kappa^2h_0}{8\big[2\alpha+\beta+2\beta\ln{(\beta R_0)}\big]
  \big[\frac{3\kappa^2h_0}{2}-\frac{\kappa(1-\gamma)}{2(1+\gamma/3)}\sqrt{2/3} V'(\xi)+2\kappa^2A^2\big]^2}\bigg]\nonumber\\
  &+&\bigg[\frac{\kappa^2h_0/\big[2\alpha+\beta+2\beta\ln{(\beta R_0)}\big]}{8\big[\frac{3\kappa^2h_0}{2}-\frac{\kappa(1-\gamma)}{2(1+\gamma/3)}\sqrt{2/3} V'(\xi)+2\kappa^2A^2\big]
  \big[\kappa^2V(\xi)+\frac{\kappa^2h_0}{2}-\frac{\kappa}{(1+\gamma/3)}\sqrt{2/3}
  V'(\xi)+\kappa^2A^2}\bigg]\, ,
  \label{D1}
\end{eqnarray}
\begin{eqnarray}
 D_2&=&2 + \epsilon_F + \epsilon_2 + \epsilon_4\nonumber\\
 &=&1 + D_1 - \epsilon_F\nonumber\\
 &=&1 + D_1 - \bigg[\frac{\frac{3\kappa^2h_0}{2} - \frac{\kappa(1-\gamma)}{2(1+\gamma/3)}\sqrt{\frac{2}{3}}V'(\xi) + 2\kappa^2A^2a_0^{2\gamma}}
  {\kappa^2A^2a_0^{2\gamma} + \kappa^2V(\xi) + \frac{\kappa^2h_0}{2} -
  \frac{\kappa}{(1+\gamma/3)}\sqrt{\frac{2}{3}}V'(\xi)}\bigg]\, ,
  \label{D2}
\end{eqnarray}
\begin{eqnarray}
 D_3&=&(1 - \epsilon_F)^2\nonumber\\
 &=&\bigg[1 - \bigg(\frac{\frac{3\kappa^2h_0}{2} - \frac{\kappa(1-\gamma)}{2(1+\gamma/3)}\sqrt{\frac{2}{3}}V'(\xi) + 2\kappa^2A^2a_0^{2\gamma}}
  {\kappa^2A^2a_0^{2\gamma} + \kappa^2V(\xi) + \frac{\kappa^2h_0}{2} -
  \frac{\kappa}{(1+\gamma/3)}\sqrt{\frac{2}{3}}V'(\xi)}\bigg)\bigg]^2\,
  ,
  \label{D3}
\end{eqnarray}
\begin{eqnarray}
 D_4&=&8\kappa^2 \frac{\rho_{KR}}{F'(R)H_F^2} + \frac{12}{F'(R)^2H_F^2}\bigg(\frac{dF'(R)}{d\tau}\bigg)^2\nonumber\\
 &=&\frac{\frac{6\kappa^2h_0\alpha}{(2+\beta/\alpha)} + 48\bigg(\frac{3\kappa^2h_0\alpha}{2} + \frac{\kappa^2A^2\alpha}{2}\bigg)^2}
 {\bigg(\frac{1}{8(1+\beta/\alpha)} + \frac{\kappa^2h_0\alpha}{2} +
 \kappa^2A^2\alpha\bigg)^2}\, .
 \label{D4}
\end{eqnarray}
From the above expressions, it is clear that the observational
indices depend on the free parameters $\alpha$, $\beta$, $h_0$,
$R_0$, $\gamma$ and $A$ where the last two arise due to the
constant-roll condition. As in the slow-roll case, we shall assume
that $\alpha \neq \beta \sim 1/R_0$, and also the parameter $A$ is
fixed in order for the expression $\kappa A\sqrt{\alpha}$ to be
equal to one, that is $\kappa A\sqrt{\alpha}=1$. In effect, the
spectral index $n_s$ and the tensor-to-scalar ratio $r$ depend
solely on the two dimensionless parameters $\kappa^h_0\alpha$,
$\beta/\alpha$, $\gamma$ and $\kappa^2A^2\alpha$. By exploring the
parameter space, we found that the observational indices are
compatible with the Planck 2018 and the BICEP2/Keck-Array data
when the values of the parameters are constrained by the bound
$\kappa^2h_0\alpha\bigg|_{max} = 5\times 10^{-5}$, which is in
agreement to the one obtained by the slow-roll approximation, and
also for $\frac{\beta}{\alpha} \simeq [0.25,0.30]$. The latter
constraint, clearly demonstrates that the ratio $\beta/\alpha$
takes more restricted values in the constant-roll case. We
summarize the results in Table[\ref{Table-2}].
\begin{table}[h]
 \centering
  \begin{tabular}{|c| c|}
   \hline \hline
   Constraints in slow-roll condition & Constraints in constant-roll condition\\
   \hline
   $\big(\frac{\kappa^2h_0}{\alpha}\big)_{max} \simeq 10^{-5}$ &
   $\big(\frac{\kappa^2h_0}{\alpha}\big)_{max} \simeq 5\times10^{-5}$ (for $\gamma = 0.01$)\\
   \hline
   $\frac{\beta}{\alpha} \simeq [0.20,0.25]$ & $\frac{\beta}{\alpha} \simeq [0.25,0.30]$ (for $\gamma = 0.01$)\\
   \hline
  \end{tabular}%
  \caption{Constraints on various parameters in slow-roll as well as in constant-roll condition}
  \label{Table-2}
 \end{table}

\begin{figure}[h]
\centering
\includegraphics[width=18pc]{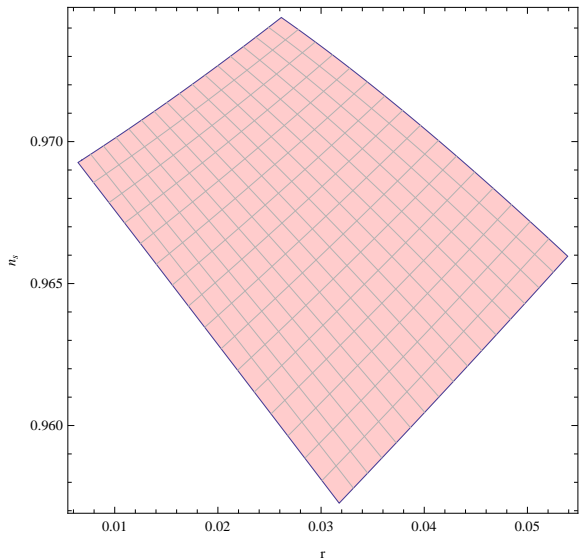}
\caption{{\it{The parametric plot of $n_s$ and $r$ as functions of
$\kappa^2h_0\alpha$ and $\beta/\alpha$, for $ 0\leq
\kappa^2h_0\alpha \leq 5\times 10^{-5}$ and $0.25\leq
\beta/\alpha\leq 0.30$.}}} \label{plot2}
\end{figure}

Therefore the present scenario successfully provides a viable
inflationary phenomenology, which is compatible with  the $Planck$
2018 and BICEP2/Keck-Array data, with the difference from the
constant-roll case being that the ratio $\beta/\alpha$ is more
constrained in the constant-roll case. We can also see the
simultaneous compatibility of the spectral index and of the
tensor-to-scalar ratio in Fig. \ref{plot2} where we present the
parametric plot of $n_s$ and $r$ as functions of
$\kappa^2h_0\alpha$ and $\beta/\alpha$, for $ 0\leq \kappa^2
h_0\leq \alpha 10^{-5}$ and $0.25\leq \beta/\alpha\leq 0.30$. As
it can be seen from Fig. \ref{plot2}, there exist a wide range of
the free parameters for which the simultaneous compatibility of
the spectral index and of the tensor-to-scalar ratio can be
achieved.

As in the slow-roll case, it is worth comparing the standard $R^2$
Starobinsky inflation phenomenology with the logarithmic $R^2$
model in the presence or not of a KR field. Recall that the
standard $R^2$ model $F(R) = R + R^2/m^2$ is viable for $0.07
\lesssim m^2/R_0 \lesssim 0.13$, with $R_0$, and it yields $0.960
\leq n_s \leq 0.970$ and $r < 0.02$. For the constant-roll case of
the $R^2$ model in the presence of a KR field, viable results are
obtained for $0.07 \lesssim m^2/R_0 \lesssim 0.13$ and
$\frac{\kappa^2h_0}{m^2} \lesssim 10^{-4}$, and particularly $n_s$
takes values in the range $0.959\leq n_s \leq 0.972$, and
$r<0.04$. For the logarithmic corrected $R^2$ model, in the
absence of the KR field, for $0.50 \lesssim m^2/R_0 \lesssim 0.60$
and $\beta = \frac{1}{2R_0}$, the spectral index $n_s$ takes
values in the range $0.958\leq n_s \leq 0.969$, and $r<0.03$.
Finally for the logarithmic corrected model in the presence of a
KR field, for $0.50 \lesssim m^2/R_0 \lesssim 0.60$,
$\frac{\kappa^2h_0}{m^2} \lesssim 5\times10^{-5}$ and $\beta =
\frac{1}{2R_0}$, $n_s$ takes values in the range $0.959\leq n_s
\leq 0.974$, and $r < 0.05$. Thus the KR field effect is to induce
a larger amount of gravitational radiation in comparison to the
standard $R^2$ model.

 \section{Conclusions}

In this paper we considered a logarithmic modification of the
standard $R^2$ Starobinsky inflation, of the form, $F(R)= R+\alpha
R^2+\beta R^2\ln{(\beta R)}$ which is motivated by one-loop
corrected higher derivative gravity, in the presence of a rank two
antisymmetric tensor field, popularly known as Kalb-Ramond field.
The model is free from ghost fields if $\alpha$, $\beta$ $> 0$.
Also the Kalb-Ramond effects are strong during the large curvature
regime, but these reduce as the Universe expands, at a rate $\sim
a^{-6}$, and in effect, radiation and matter dominate the
post-inflationary phase of our Universe. We focused on the
inflationary aspects of the model, and we calculated in detail the
observational indices of the model, in two approximating cases,
namely under the slow-roll condition and under the constant-roll
condition. It turns out that for both the slow-roll and
constant-roll cases, the theoretical values of $n_s$, $r$ match
the observational constraints if the values of $h_0$ and $\alpha$
take values that are bounded from above by the constraint
$\kappa^2h_0\alpha\bigg|_{max} = 5\times 10^{-5}$, where $h_0$
denotes the energy density of the Kalb-Ramond field. However the
ratio $\beta/\alpha$ is constrained in different ways, in the
slow-roll and constant-roll cases, with the constant-roll
constraint being $0.25 < \frac{\beta}{\alpha} \lesssim 0.30$ and
the slow-roll one being $0.20 < \frac{\beta}{\alpha} \lesssim
0.25$. In addition, our theoretical framework puts an upper bound
in the value of $h_0$, which can be obtained by fixing $\alpha =
1$ in Planck units, and this leads to $h_0^{max} \sim 10^{71}$
(GeV)$^{4}$, so in effect, the Kalb-Ramond field energy density is
constrained as follows $\big(\rho_{KR}\big)^{max} \sim 10^{71}$
(GeV)$^{4}$. Moreover, let us note that the effect of the KR field
on the inflationary phenomenology of $R^2$ Starobinsky inflation
and logarithmic $R^2$ gravity is that it increases the amount of
the primordial gravitational radiation. As a generalization of
this work, one should consider the effects of a massive
antisymmetric field \cite{Buchbinder:2008jf} on $F(R)$ gravity
inflation, and we hope to address this issue in a future work.

Finally we need to stress an important issue, the fact that in the
context of our work, there is a direct correspondence between the
massless KR field and the massless scalar field at the level of
cosmological background and linear cosmological perturbations
(with the latter perturbations equivalence being considered with
caution and only at linear order). Particularly, the
correspondence is materialized by the equivalence of the Eqs.
(\ref{ref_perturbed_eqn1} - \ref{ref_perturbed_eqn7}) and Eqs.
(\ref{ref_perturbed_eqn8} - \ref{ref_perturbed_eqn14}), which can
be used as a recipe to go between the two pictures. This
correspondence recipe, at this level of the analysis, allows to
obtain all the results for KR fields in inflation directly from
the massless scalar results which are considerably simpler to
calculate. Indeed the equivalence is justified due to the fact
that, $H_{\mu\nu\alpha} =
\epsilon_{\mu\nu\rho\sigma}\partial^{\sigma}Z$ (with
$H_{\mu\nu\alpha}$ and $Z$ being the KR tensor and the scalar
field respectively) which can be indeed used as a recipe to go
between the two pictures. Thereby this correspondence relation
allows one to obtain the expression of inflationary parameters
with KR field directly from massless scalar case, which are
relatively simpler to compute, as we already mentioned.
Furthermore, in the present context as we are interested on
inflationary parameters like spectral index and tensor to scalar
ratio, so we demonstrated the equivalence only up to first order
perturbation equations and we did not consider the higher order
perturbations. However the investigation of such equivalence for
second (or higher) order perturbations is important from its own
right, and non-trivial, so we hope to address in a future work.
Also, we should note that we studied free KB tensors, but in
principle we can  also take into account potential terms for them,
and /or we can consider a direct non-minimal coupling of such
fields with curvature for example of the form $G(F^2) f(R)$, where
$F^2$ is the square of such tensors. In this case, the background
equivalence is in general lost and the results would completely
different from the case we studied in this paper, in which only
non-minimal couplings were considered. This extension will be
considered elsewhere.

\section*{Acknowledgments}

This work is supported by MINECO (Spain), FIS2016-76363-P and by
project 2017 SGR247 (AGAUR, Catalonia) (E.E and S.D.O).  T. Paul
sincerely thanks to the institute ICE-CSIC, Spain where the work
has been done, for their warm hospitality.

 \section*{Appendix: The Time Dependence of the Kalb-Ramond Field and Equivalence of Field Equations}

Due to antisymmetric nature, $\tilde{H}_{\mu\nu\alpha}$ has four
independent components in four dimensions and thus it can be
equivalently expressed as a vector field as follows,
\begin{eqnarray}
 \tilde{H}_{\mu\nu\alpha} =
 \varepsilon_{\mu\nu\alpha\beta}\Upsilon^{\beta}\, ,
 \label{app1 1}
\end{eqnarray}
where $\varepsilon_{\mu\nu\alpha\beta}$ is the Levi-Civita symbol
and $\Upsilon^{\beta}$ is a vector field propagating in four
dimensional spacetime. The four components of $\Upsilon^{\beta}$
are connected with the independent components of
$\tilde{H}_{\mu\nu\alpha}$ as follows,
\begin{eqnarray}
\tilde{H}_{012} = h_1 = \Upsilon^{3},\,\,\,\,\,\,\tilde{H}_{013} = h_2 = -\Upsilon^2\nonumber\\
\tilde{H}_{023} = h_3 = \Upsilon^{1},\,\,\,\,\,\,\tilde{H}_{123} =
h_4 = -\Upsilon^0\, . \label{app1 2}
 \end{eqnarray}
By using a FRW background, the off-diagonal Friedmann equations
become,
 \begin{eqnarray}
  \Upsilon_3\Upsilon^2 = \Upsilon_3\Upsilon^1 = \Upsilon_2\Upsilon^1 = \Upsilon_0\Upsilon^3 = \Upsilon_0\Upsilon^2 = \Upsilon_0\Upsilon^1 =
  0\, .
  \label{app1 3}
 \end{eqnarray}
The above set of equations clearly indicate that only one
component of $\Upsilon^{\beta}$ is non-zero, which in turn reduces
the independent components of $\tilde{H}_{\mu\nu\alpha}$ to one.
Therefore, in the present context (i.e. for spatially flat FRW
metric in four dimensions), $\Upsilon^{\beta}$ can be expressed as
a derivative of a massless scalar field $Z(x^{\mu})$ (i.e
$\Upsilon^{\beta} = \partial^{\beta}Z$),  which further relates
the KR field tensor with the scalar field in the following way,
 \begin{eqnarray}
  \tilde{H}_{\mu\nu\alpha}&=&\varepsilon_{\mu\nu\alpha\beta}\Upsilon^{\beta}\nonumber\\
  &=&\varepsilon_{\mu\nu\alpha\beta}\partial^{\beta}Z
  \label{app1 4}
 \end{eqnarray}
 For the FRW metric, the scalar field ($Z$) is  spatially homogeneous and in effect, its equation of motion turns
 out to be,
 \begin{eqnarray}
  \ddot{Z} + 3H\dot{Z} = 0\, .
  \label{app1 5}
 \end{eqnarray}
 where $H$ is the Hubble parameter. Solving the above equation, one
 obtains,
 \begin{eqnarray}
  \frac{\partial Z}{\partial t} \propto \frac{1}{a^3} = \frac{d}{a^3}
  \label{app1 6}
 \end{eqnarray}
 where $d$ is an integration constant. With this solution of $\frac{\partial Z}{\partial t}$, the
diagonal Friedmann equations take the following form,
 \begin{eqnarray}
 H^2&=&\frac{\kappa^2}{3}\bigg[\frac{1}{2}\dot{\xi}^2 + \frac{m^2}{8\kappa^2}\big(1 - e^{\sqrt{\frac{2}{3}}\kappa\xi}\big)^2
 + \frac{1}{2}\dot{Z}^2\bigg]\nonumber\\
 &=&\frac{\kappa^2}{3}\bigg[\frac{1}{2}\dot{\xi}^2 + \frac{m^2}{8\kappa^2}\big(1 - e^{\sqrt{\frac{2}{3}}\kappa\xi}\big)^2
 + \frac{d}{2a^6}\bigg]\, ,
 \label{app1 7}
\end{eqnarray}
and
\begin{eqnarray}
 2\dot{H} + 3H^2 = -\kappa^2\bigg[\frac{1}{2}\dot{\xi}^2
 - \frac{m^2}{8\kappa^2}\bigg(1 - e^{\sqrt{\frac{2}{3}}\kappa\xi}\bigg)^2 + \frac{1}{2}\dot{Z}^2\bigg]\nonumber\\
 = -\kappa^2\bigg[\frac{1}{2}\dot{\xi}^2
 - \frac{m^2}{8\kappa^2}\bigg(1 - e^{\sqrt{\frac{2}{3}}\kappa\xi}\bigg)^2 +
 \frac{d}{2a^6}\bigg]\, ,
 \label{app1 8}
\end{eqnarray}
and recall, $\xi(t)$ is the scalar field arises from the higher
curvature degree of freedom. Furthermore, the field equation for
$\xi(t)$ is given by,
\begin{eqnarray}
 \ddot{\xi} + 3H\dot{\xi} - \sqrt{\frac{2}{3}}\frac{m^2}{4\kappa}e^{\sqrt{\frac{2}{3}}\kappa\xi}\big(1 - e^{\sqrt{\frac{2}{3}}\kappa\xi}\big) =
 0\, .
 \label{app1 9}
\end{eqnarray}
As it can be seen the above equations match the field equations
appearing in Eqs. (\ref{independent equation1}), (\ref{independent
equation2}), by identifying the constant $d$ with $h_0$. This
leads to the argument that the field equations of the KR field
obtained with and without expressing the KR as a vector field, are
equivalent.

\end{document}